\documentclass[11pt,a4paper]{article}
\usepackage[a4paper,left=24mm,right=24mm,top=24mm,bottom=25mm,headheight=15pt]{geometry}
\usepackage{amsmath,amssymb,amsthm,mathtools,bm,mathrsfs}
\usepackage{fontspec}
\usepackage{graphicx,booktabs,array,tabularx,longtable}
\usepackage{microtype,caption,subcaption,fancyhdr,placeins,needspace,etoolbox}
\usepackage[numbers,sort&compress]{natbib}
\usepackage[hidelinks,bookmarksnumbered]{hyperref}
\usepackage{xurl}
\newtheorem{theorem}{Theorem}
\newtheorem{lemma}{Lemma}
\newtheorem{proposition}{Proposition}
\theoremstyle{remark}
\numberwithin{equation}{section}
\allowdisplaybreaks[2]
\pretocmd{\section}{\Needspace{6\baselineskip}}{}{}
\pretocmd{\subsection}{\Needspace{5\baselineskip}}{}{}
\newcommand{\dd}{\mathrm{d}}
\newcommand{\vv}{\boldsymbol v}
\newcommand{\BB}{\boldsymbol B}
\newcommand{\JJ}{\boldsymbol J}
\newcommand{\Avec}{\boldsymbol A}
\newcommand{\EE}{\boldsymbol E}
\newcommand{\MM}{\boldsymbol M}
\newcommand{\xx}{\boldsymbol x}
\newcommand{\XX}{\boldsymbol X}
\newcommand{\nn}{\boldsymbol n}
\newcommand{\xii}{\boldsymbol\xi}
\newcommand{\atest}{\boldsymbol a}
\newcommand{\ttau}{\boldsymbol\tau}
\newcommand{\etab}{\boldsymbol\eta}
\newcommand{\Id}{\boldsymbol I}
\newcommand{\qv}{\boldsymbol q}
\newcommand{\grad}{\nabla}
\newcommand{\diver}{\nabla\!\cdot}
\newcommand{\curl}{\nabla\!\times}
\newcommand{\Dt}{\mathrm D_t}
\newcommand{\Cv}{\mathcal C_{\vv}}
\newcommand{\st}{\mathfrak s}
\newcommand{\sym}{\operatorname{sym}}
\newcommand{\skw}{\operatorname{skw}}
\newcommand{\sech}{\operatorname{sech}}

\usepackage{float}
\newcommand{\repositoryurl}{\url{https://github.com/hl-ren/pdel-lean-verification}}
\newcommand{\repositoryarchive}{\href{https://github.com/hl-ren/pdel-lean-verification/releases/tag/v1.0.1}{verification release v1.0.1}}

\hypersetup{pdftitle={A path-dependent entropic Lagrangian formulation for exact thermoviscous evolution of sheared magnetic tori with anisotropic resistivity},pdfauthor={Huilong Ren}}
\title{\bfseries A path-dependent entropic Lagrangian formulation for exact thermoviscous evolution of sheared magnetic tori with anisotropic resistivity}
\author{Huilong Ren\\[0.3em]
\small College of Civil Engineering, Tongji University, Shanghai 200092, China\\
\small \texttt{hlren@tongji.edu.cn}}
\date{}
\begin{document}
\maketitle
\begin{abstract}
Resistive evolution of a three-dimensional magnetostatic equilibrium requires compatibility among induction, momentum balance and thermal transport. We construct exact thermoviscous evolutions of a non-axisymmetric magnetic torus with nonzero magnetic shear using a path-dependent entropic Lagrangian (PDEL) formulation. A symmetric positive-definite resistivity tensor is constructed so that the resistive differential expression preserves the spatial form of a known sheared toroidal equilibrium. Homogeneous expansion is coupled to an ideal gas and positive bulk viscosity in the absence of body forces and volumetric heat supply. For a one-temperature gas with heat-capacity ratio $5/3$, a $T^{-2/3}$ resistivity prefactor gives elementary time dependence for all fields. A two-temperature extension admits a $T_e^{-3/2}$ prefactor and positive interspecies heat exchange on a finite state interval. Material magnetic flux decays while the rotational-transform profile is preserved. Both branches satisfy local total-energy conservation, moving-boundary power balance and material transport of dissipation histories. A seed-specific obstruction excludes positive scalar resistivity along the same homologous trajectory; the two-temperature exchange law has a nonremovable endpoint pole. Symbolic identities, independent Cartesian differentiation and surface--volume quadrature verify these results. The solutions provide coupled MHD benchmarks with passive, inverse-designed constitutive coefficients; those coefficients are not identified with a classical collisional plasma closure.
\end{abstract}
\section{Introduction}\label{sec:intro}
Magnetohydrodynamic (MHD) evolution couples magnetic geometry to force balance, flux transport and thermal energy. Classical toroidal equilibrium theory describes magnetic surfaces, rotational transform and the constraints imposed by three-dimensional geometry \citep{Kruskal1958,Grad1967,Helander2014}. Extending an equilibrium to a dissipative evolution also requires resistive flux loss, irreversible heating and boundary power to satisfy the thermal and mechanical balances. In a compressible medium, instantaneous force balance alone does not ensure compatibility with the equation of state and the temperature equation.

Analytic equilibria provide reference geometries for this problem. Landreman \citep{Landreman} constructed explicit Cartesian non-axisymmetric equilibria with nested toroidal surfaces and nonconstant rotational transform. Issan et al.\ \citep{Issan} extended these families to configurations without stellarator symmetry. G\'omez-Serrano et al.\ \citep{Gomez2026} established smooth equilibria on embedded solid tori with prescribed discrete symmetry groups by a different construction. Exact shearing-box solutions with viscosity, resistivity and cooling, together with nonlinear helical-wave solutions, illustrate compatibility conditions for dissipative dynamics \citep{Lesaffre2007,Zhu2014}. The magnetic--thermal compatibility of resistively diffusing toroidal equilibria was examined by Montani et al.\ \citep{Montani}. The present work seeks an explicit dissipative evolution that retains a three-dimensional toroidal geometry and satisfies the coupled balance equations.

The constitutive specification determines which such evolutions are possible. Spitzer--H\"arm and Braginskii transport theory derives electrical and thermal coefficients from collisional plasma physics \citep{Spitzer,Braginskii1965}. Anisotropic resistivity has also been considered in axisymmetric equilibria with toroidal flow, and nonlinear resistive laws have been derived from collisional momentum transfer \citep{Poulipoulis,Lingam}. Here the coefficients are instead constructed for a prescribed continuum trajectory: the resistivity tensor, bulk-viscosity texture and interspecies exchange law are determined from reference fields and transported with the material. The construction enforces passivity and thermal compatibility without identifying these coefficients with a classical collisional closure.

Variational methods express the coupling among material motion, field storage and irreversible power. Newcomb's MHD action and Euler--Poincar\'e reduction relate advected quantities to Eulerian variations in the reversible setting \citep{Newcomb1962,Holm1998}. Nonequilibrium variational thermodynamics, port-Hamiltonian formulations and metriplectic descriptions incorporate dissipation and boundary exchange \citep{GayBalmaz2017a,GayBalmaz2017b,PortHamiltonian,Materassi2012,Coquinot2020}. For MHD, Carlier and Campos-Pinto \citep{Carlier} developed a visco-resistive Hall formulation, while structure-preserving finite element methods enforce magnetic constraints and discrete energy or entropy balances \citep{Gawlik2021,Gawlik2025}. Exact solutions with independently evaluable thermal and boundary powers provide benchmarks for these coupled formulations.

The path-dependent entropic Lagrangian (PDEL) formulation introduces accumulated process powers into a state--history functional \citep{RenEntropy,Ren}. Its terminal differential distinguishes arbitrary weak variations from evaluation of a constitutive action on realized rates; the endpoint and weak power-port structure is developed in \citet{RenSpending}. Eulerian PDEL treats thermoviscous flow with species diffusion and reaction \citep{RenEulerian}, and its multi-channel extension describes simultaneous power conversion among irreversible processes \citep{RenMultiChannel}. These constructions specify the material-to-spatial transport of histories and the pairing of donor and receiver powers. In the MHD model considered here, magnetic energy is retained in the functional, and one resistive constitutive action supplies the vector Ohm equation, mechanical Lorentz action and Joule receiver.

A strictly positive current--potential pairing permits a symmetric positive-definite tensor to map the reference current to its vector potential. The curl of the resulting resistive field is proportional to the reference magnetic field. Homogeneous expansion then gives explicit time dependence when density, viscosity and temperature are chosen to satisfy momentum and thermal balance. The resulting one-temperature branch has a $T^{-2/3}$ resistivity prefactor, and a finite-interval two-temperature branch has a $T_e^{-3/2}$ prefactor. Both exhibit material magnetic-flux loss with an unchanged rotational-transform profile. A local scalar-resistivity obstruction and a nonremovable exchange endpoint delimit the construction. Parameter studies characterize the geometry, time scale and heat partition; stability and confinement performance are outside the analysis.

The remainder of the paper is outlined as follows. Section~\ref{sec:review} introduces the PDEL operations through a thermo-viscoelastic example and its Eulerian and magnetic extensions. Section~\ref{sec:pdel} gives the MHD functional and governing equations, and Section~\ref{sec:realization} constructs the passive resistive tensor for the toroidal seed. The one- and two-temperature solutions are derived in Sections~\ref{sec:oneT} and \ref{sec:twoT}, with energy and history identities in Section~\ref{sec:energy}. Parameter studies and verification are presented in Section~\ref{sec:verification}, followed by conclusions in Section~\ref{sec:discussion}. Appendices~\ref{app:geometry}--\ref{app:review} contain the geometric calculations, constitutive restrictions and first variation of the introductory continuum example.


\section{Review of the PDEL construction}\label{sec:review}
\subsection{A thermo-viscoelastic continuum and its explicit power histories}
The small-strain thermo-viscoelastic example fixes the state and history operations before their Eulerian extension \citep{RenEntropy,RenSpending,RenEulerian,RenMultiChannel}. Let $\Omega_0$ be a fixed reference body, $\boldsymbol u(\XX,t)$ its displacement, $\theta>0$ its temperature, and $s$ its entropy per reference volume. Set $\boldsymbol\varepsilon=\sym\grad_X\boldsymbol u$ and use the reference-volume Helmholtz density
\begin{equation}
\begin{aligned}
\psi(\boldsymbol\varepsilon,\theta)
={}&\tfrac12\boldsymbol\varepsilon:\mathbb C:\boldsymbol\varepsilon
-(\theta-\theta_0)\boldsymbol\beta_T:\boldsymbol\varepsilon\\
&+\rho_0c_v\left[(\theta-\theta_0)-\theta\log(\theta/\theta_0)\right].
\end{aligned}
\label{eq:reviewfree}
\end{equation}
Here $\rho_0c_v>0$ is the reference volumetric heat capacity, $\mathbb C$ is the elastic tensor, and $\boldsymbol\beta_T$ is the thermoelastic stress-coupling tensor. These coefficients are independent of temperature in this example. The reversible stress, Kelvin--Voigt stress and referential heat flux are
\begin{equation}
\begin{aligned}
\boldsymbol\sigma_e&=\psi_{,\boldsymbol\varepsilon}
=\mathbb C:\boldsymbol\varepsilon-(\theta-\theta_0)\boldsymbol\beta_T,\\
\boldsymbol\sigma_v&=\mathbb V:\dot{\boldsymbol\varepsilon},\qquad
D_v=\boldsymbol\sigma_v:\dot{\boldsymbol\varepsilon}
=\dot{\boldsymbol\varepsilon}:\mathbb V:\dot{\boldsymbol\varepsilon}\ge0,\\
\boldsymbol Q&=-\boldsymbol K_0\grad_X\theta,
\qquad \boldsymbol K_0=\boldsymbol K_0^T\succeq0.
\end{aligned}
\label{eq:reviewconstitutive}
\end{equation}
The viscosity tensor $\mathbb V$ is positive semidefinite on symmetric strain rates. With no volumetric heat supply, the two history densities have the explicit integrands
\begin{equation}
\begin{aligned}
H_Q(\XX,t)&=\int_{t_0}^{t}\grad_X\!\cdot\boldsymbol Q(\XX,\tau)\dd\tau
=-\int_{t_0}^{t}\grad_X\!\cdot(\boldsymbol K_0\grad_X\theta)(\XX,\tau)\dd\tau,\\
H_{D_v}(\XX,t)&=\int_{t_0}^{t}
\boldsymbol\sigma_v(\XX,\tau):\dot{\boldsymbol\varepsilon}(\XX,\tau)\dd\tau.
\end{aligned}
\label{eq:reviewhistories}
\end{equation}
Thus $\boldsymbol Q$ is a heat-flux vector, $\grad_X\!\cdot\boldsymbol Q$ is signed outward thermal power per reference volume, and $D_v$ is local irreversible conversion. For prescribed body-force density $\boldsymbol b_0$ and traction $\overline{\boldsymbol t}$, the state--history functional is
\begin{equation}
\begin{aligned}
\mathscr A_{\rm tv}={}&\int_{t_0}^{t_1}\!\int_{\Omega_0}
\left[\tfrac12\rho_0|\dot{\boldsymbol u}|^2-\psi-\theta s
-H_Q+H_{D_v}+\boldsymbol b_0\cdot\boldsymbol u\right]\dd V_0\dd t\\
&+\int_{t_0}^{t_1}\!\int_{\Gamma_t}
\overline{\boldsymbol t}\cdot\boldsymbol u\,\dd S_0\dd t.
\end{aligned}
\label{eq:reviewaction}
\end{equation}
Entropy and temperature remain independent mixed-state variables until the thermal state variation is taken. The histories organize realized power exchange; they are not additional physical stored energies.

\subsection{Independent variations and continuum field equations}
Keep an arbitrary compatible displacement variation $\delta\boldsymbol u$, with
$\delta\boldsymbol\varepsilon=\sym\grad_X\delta\boldsymbol u$, and an independent temperature variation $\delta\theta$. The displacement variation vanishes at the time endpoints and on the prescribed-displacement boundary. The independent entropy selector is $\delta t_s$, with $\delta^{\rm ch}s=\dot s\,\delta t_s$. The locked-history operations are
\begin{equation}
\begin{aligned}
\delta^{\rm ch}H_Q&=(\grad_X\!\cdot\boldsymbol Q)\,\delta t_s,\\
\delta^{\rm ch}H_{D_v}
&=D_v\,\delta t_s-\boldsymbol\sigma_v:\delta\boldsymbol\varepsilon.
\end{aligned}
\label{eq:reviewrules}
\end{equation}
The receiver term is the terminal derivative of the committed integral; the donor term is the weak action of the same constitutive stress \citep{RenSpending,RenMultiChannel}. On the realized mechanical strain direction,
\begin{equation}
\delta^{\rm ch}\boldsymbol\varepsilon
=\dot{\boldsymbol\varepsilon}\,\delta t_m,
\qquad
\boldsymbol\sigma_v:\delta^{\rm ch}\boldsymbol\varepsilon
=D_v\,\delta t_m,
\qquad
\delta^{\rm ch}H_{D_v}=D_v(\delta t_s-\delta t_m).
\label{eq:reviewcalibration}
\end{equation}
This is a calibration of the power pairing, not a restriction of the arbitrary mechanical test. In particular, for spatially varying $\delta t_m$, the ordinary strain variation is formed before this evaluation; it is not replaced by $\sym\grad_X(\dot{\boldsymbol u}\,\delta t_m)$.

Applying the state and channel operations to \eqref{eq:reviewaction} and integrating by parts gives
\begin{equation}
\begin{aligned}
\mathfrak D\mathscr A_{\rm tv}={}&-\int_{t_0}^{t_1}\!\int_{\Omega_0}
\Big\{\big[\rho_0\ddot{\boldsymbol u}
-\grad_X\!\cdot(\boldsymbol\sigma_e+\boldsymbol\sigma_v)-\boldsymbol b_0\big]
\cdot\delta\boldsymbol u\\
&\qquad +(\psi_{,\theta}+s)\delta\theta
+\big[\theta\dot s+\grad_X\!\cdot\boldsymbol Q-D_v\big]\delta t_s\Big\}\dd V_0\dd t\\
&+\int_{t_0}^{t_1}\!\int_{\Gamma_t}
\big[\overline{\boldsymbol t}-(\boldsymbol\sigma_e+\boldsymbol\sigma_v)\boldsymbol N\big]
\cdot\delta\boldsymbol u\,\dd S_0\dd t.
\end{aligned}
\label{eq:reviewfirstvariation}
\end{equation}
The independent variations recover
\begin{equation}
\begin{aligned}
\rho_0\ddot{\boldsymbol u}
&=\grad_X\!\cdot(\boldsymbol\sigma_e+\boldsymbol\sigma_v)+\boldsymbol b_0,\\
s=-\psi_{,\theta}
&=\boldsymbol\beta_T:\boldsymbol\varepsilon+\rho_0c_v\log(\theta/\theta_0),\\
\theta\dot s+\grad_X\!\cdot\boldsymbol Q&=D_v,
\qquad (\boldsymbol\sigma_e+\boldsymbol\sigma_v)\boldsymbol N
=\overline{\boldsymbol t}\quad\text{on }\Gamma_t.
\end{aligned}
\label{eq:reviewequations}
\end{equation}
Consequently,
\begin{equation}
\rho_0c_v\dot\theta
+\theta\boldsymbol\beta_T:\dot{\boldsymbol\varepsilon}
-\grad_X\!\cdot(\boldsymbol K_0\grad_X\theta)
=\dot{\boldsymbol\varepsilon}:\mathbb V:\dot{\boldsymbol\varepsilon}.
\label{eq:reviewtemperature}
\end{equation}
For $e_0=\psi+\theta s$, the same stress power cancels between mechanical and internal energy:
\begin{equation}
\partial_t\left(\tfrac12\rho_0|\dot{\boldsymbol u}|^2+e_0\right)
+\grad_X\!\cdot\left[\boldsymbol Q
-(\boldsymbol\sigma_e+\boldsymbol\sigma_v)^T\dot{\boldsymbol u}\right]
=\boldsymbol b_0\cdot\dot{\boldsymbol u}.
\label{eq:reviewenergy}
\end{equation}
Appendix~\ref{app:review} gives the pre-integration-by-parts expression, boundary budget and entropy-flux form. Related nonequilibrium variational theories also organize stress power and entropy \citep{GayBalmaz2017a,GayBalmaz2017b}; here their association is specified by the PDEL terminal-history and weak-power rules.

The Lean formalization of this example is restricted to a homogeneous single-degree-of-freedom reduction. The spatial integration by parts and boundary traction in \eqref{eq:reviewfirstvariation} lie outside that reduction. The formalized Eulerian MHD specialization is described in Section~\ref{sec:formal}.

\subsection{Eulerian transport and multi-channel power conversion}
The Eulerian PDEL formulation extends the preceding material construction to thermoviscous flow with species transport and reaction \citep{RenEulerian}. Two transport operators are needed from the outset. For a scalar field $f$, the material derivative follows its value along a moving parcel; for a density $w$ measured per current volume, the conservative rate also accounts for local volume change:
\begin{equation}
\Dt f:=\partial_t f+\vv\cdot\grad f,\qquad
\Cv w:=\partial_t w+\diver(w\vv)
=\Dt w+w\diver\vv.
\label{eq:reviewrates}
\end{equation}
The material derivative acts componentwise on vectors. The calligraphic $\Cv$ is a transport operator, distinct from the specific heat $c_v$. Thus $T$ is differentiated by $\Dt$, whereas $\rho$, $\st$ and the current-volume power histories use $\Cv$. For example, $\Cv\rho=0$ expresses mass conservation; it does not imply $\Dt\rho=0$ in a compressible flow.

The channel labels identify the participating physical processes. The mechanical channel $m$ carries stress work, the entropy channel $s$ receives the associated irreversible power, and the electromagnetic channel $e$ carries field work when the field is retained. Their realized-path selectors are $\delta t_m$, $\delta t_s$ and $\delta t_e$. These selectors calibrate the respective powers; they do not replace arbitrary displacement, temperature or vector-potential variations. In a two-temperature model, independent thermal selectors $\delta t_{s_e}$ and $\delta t_{s_i}$ distinguish the electron and ion receivers. Species channels are not needed in the nonreacting specialization considered here.

For the single-temperature fluid without a magnetic field, the state--history functional is
\begin{equation}
\mathscr A_E=\int\!\int_{\Omega(t)}
\left[\tfrac12\rho|\vv|^2-\Phi(\rho,T)-T\st
-H_{\mathrm{div}\,q}+H_\nu\right]\dd V\dd t.
\label{eq:revieweuleraction}
\end{equation}
The explicit histories follow the material parcel $\xx=\xx(\XX,t)$, with Jacobian $\mathcal J>0$:
\begin{equation}
\begin{aligned}
H_{\mathrm{div}\,q}(\xx(\XX,t),t)
&=\frac1{\mathcal J(\XX,t)}\int_{t_0}^{t}
\mathcal J(\XX,\tau)\,\diver\qv(\xx(\XX,\tau),\tau)\dd\tau,\\
H_\nu(\xx(\XX,t),t)
&=\frac1{\mathcal J(\XX,t)}\int_{t_0}^{t}
\mathcal J(\XX,\tau)\,[\ttau:\grad\vv](\xx(\XX,\tau),\tau)\dd\tau.
\end{aligned}
\label{eq:revieweulerhistories}
\end{equation}
More generally, the push-forward of one current-volume power density $g$ is
\begin{equation}
H_g(\xx(\XX,t),t)
=\frac1{\mathcal J(\XX,t)}\int_{t_0}^{t}
\mathcal J(\XX,\tau)g(\xx(\XX,\tau),\tau)\dd\tau,
\qquad \mathcal C_{\vv}H_g=g.
\label{eq:reviewtransport}
\end{equation}
The constitutive choices are $\qv=-\boldsymbol K\grad T$ and, in the present bulk-viscous model, $\ttau=\zeta(\diver\vv)\Id$; hence $D_\nu=\ttau:\grad\vv=\zeta(\diver\vv)^2$. To display how the equations follow from \eqref{eq:revieweuleraction}, retain an arbitrary Eulerian displacement $\xii$ and the standard relabeling directions \citep{Holm1998,RenEulerian}:
\begin{equation}
\begin{aligned}
\delta_\xi\vv&=\partial_t\xii+(\vv\cdot\grad)\xii-(\xii\cdot\grad)\vv,&
\delta_\xi\rho&=-\diver(\rho\xii),\\
\delta_\xi T&=-\xii\cdot\grad T,&
\delta_\xi\st&=-\diver(\st\xii).
\end{aligned}
\label{eq:revieweulervariations}
\end{equation}
The temperature test $\delta_T T=\delta T$ is independent of this relabeling. The entropy channel has $\delta^{\rm ch}\st=\Cv\st\,\delta t_s$, and the two history operations are
\begin{equation}
\delta^{\rm ch}H_{\mathrm{div}\,q}=(\diver\qv)\delta t_s,
\qquad
\delta^{\rm ch}H_\nu=D_\nu\delta t_s-\ttau:\grad\xii.
\label{eq:revieweulerspending}
\end{equation}
The last term is the mechanical donor and the first term in the same viscous rule is its entropy receiver. Evaluating the already formed stress pairing on the physical rate gives $D_\nu\delta t_m$; no replacement of the arbitrary field $\xii$ is made inside a spatial gradient.

For compactly supported spatial tests and vanishing temporal endpoint displacement, integration by parts uses $\Cv\rho=0$ and
\[
\delta_\xi\!\int\!\int\tfrac12\rho|\vv|^2\dd V\dd t
=-\int\!\int\rho\Dt\vv\cdot\xii\,\dd V\dd t,
\qquad
-\rho\grad\Phi_\rho+\Phi_T\grad T=-\grad p,
\quad p:=\rho\Phi_\rho-\Phi.
\]
The pressure identity follows by differentiating $p$; the mixed $T\st$ terms cancel in the mechanical coefficient without first imposing thermal conjugacy. Adding the history terms yields the bulk first differential
\begin{equation}
\begin{aligned}
\mathfrak D\mathscr A_E=\int\!\int_{\Omega(t)}\Big\{&
[-\rho\Dt\vv-\grad p+\diver\ttau]\cdot\xii
-(\Phi_T+\st)\delta T\\
&+[-T\Cv\st-\diver\qv+D_\nu]\delta t_s\Big\}\dd V\dd t.
\end{aligned}
\label{eq:revieweulerfirstvariation}
\end{equation}
Since $\xii$, $\delta T$ and $\delta t_s$ are independent, stationarity gives
\begin{equation}
\rho\Dt\vv=-\grad p+\diver\ttau,\qquad
\st=-\Phi_T,\qquad
T\Cv\st+\diver\qv=D_\nu.
\label{eq:revieweulerbalances}
\end{equation}
The mechanical, conjugacy and entropy equations therefore arise as three coefficients of the same differential. Section~\ref{sec:pdel} retains these operations and adds independent magnetic storage and its resistive power pairing.

The submitted multi-channel formulation \citep{RenMultiChannel} addresses the additional case in which one realized power is transferred between several processes. For an energy-transfer rate $Q_{i\to e}$, its history contributes $Q_{i\to e}(\delta t_{s_e}-\delta t_{s_i})$: the receiving and donating thermal channels use the same quantity with opposite signs. Its total-energy contribution cancels, while its entropy contribution is $Q_{i\to e}(T_e^{-1}-T_i^{-1})$. This is the multi-channel structure used in Section~\ref{sec:twoT}; it does not prescribe a microscopic exchange coefficient.

\subsection{Retained electromagnetic storage and explicit Joule history}
A fixed conducting medium provides a retained-electromagnetic-field specialization of the same common-power construction. To identify the storage signs and history integrands explicitly, choose a compatible temporal-gauge potential with $\EE=-\partial_t\Avec$, $\BB=\curl\Avec$, constant $\epsilon_0,\mu_0,\sigma_c>0$, and conduction current $\JJ_c=\sigma_c\EE$. With $\qv=-\boldsymbol K\grad T$, set
\begin{equation}
\begin{aligned}
H_q(\xx,t)&=\int_{t_0}^{t}\diver\qv(\xx,\tau)\dd\tau,\\
H_J(\xx,t)&=\int_{t_0}^{t}\JJ_c(\xx,\tau)\cdot\EE(\xx,\tau)\dd\tau
=\int_{t_0}^{t}\sigma_c|\EE(\xx,\tau)|^2\dd\tau.
\end{aligned}
\label{eq:reviewEMhistories}
\end{equation}
The histories are ordinary fixed-position integrals only because the carrier medium is stationary. The corresponding functional is
\begin{equation}
\mathscr A_{\rm EM}=\int\!\int_\Omega
\left[\tfrac12\epsilon_0|\EE|^2-\tfrac1{2\mu_0}|\BB|^2
-\psi_{\rm th}(T)-T\st-H_q+H_J\right]\dd V\dd t,
\label{eq:reviewmaxwellaction}
\end{equation}
with the explicitly calibrated operations
\begin{equation}
\begin{aligned}
\delta^{\rm ch}H_q&=(\diver\qv)\delta t_s,\\
\delta^{\rm ch}H_J&=\sigma_c|\EE|^2\delta t_s+\JJ_c\cdot\delta\Avec,
\qquad \delta^{\rm ch}\st=\partial_t\st\,\delta t_s.
\end{aligned}
\label{eq:reviewEMrules}
\end{equation}
On the realized electromagnetic direction $\delta^{\rm ch}\Avec=\partial_t\Avec\,\delta t_e=-\EE\delta t_e$, the last weak term is $-D_J\delta t_e$. In the variational derivation, however, $\delta\Avec$ remains an arbitrary independent vector field. Its coefficient gives $\curl\boldsymbol H=\epsilon_0\partial_t\EE+\JJ_c$, with $\boldsymbol H=\BB/\mu_0$, while $\delta T$ and $\delta t_s$ give $\st=-\partial_T\psi_{\rm th}$ and $T\partial_t\st+\diver\qv=\sigma_c|\EE|^2$. Faraday's law and $\diver\BB=0$ follow from the potential representation; compatible initially neutral Gauss data are preserved for these constant coefficients.

Retained field storage and the electromagnetic boundary flux satisfy
\begin{equation}
\partial_t\left(\frac{\epsilon_0|\EE|^2}{2}
+\frac{|\BB|^2}{2\mu_0}\right)
+\diver(\EE\times\boldsymbol H)=-\JJ_c\cdot\EE.
\label{eq:reviewpoynting}
\end{equation}
The distinction between field storage, boundary power and internal conversion is consistent with port-based and dissipative variational descriptions \citep{PortHamiltonian,Carlier,Gawlik2025}. The magnetic-quasistatic limit used below neglects electric field storage but retains magnetic storage and an independent vector test. For a moving medium,
\begin{equation}
\JJ\cdot\EE=\vv\cdot(\JJ\times\BB)
+\JJ\cdot(\EE+\vv\times\BB).
\label{eq:reviewEMsplit}
\end{equation}
The relative electric-field power is the resistive receiver after constitutive closure; the Lorentz term is field--motion exchange. This separation, rather than treating all electromagnetic work as heat, connects the continuum and multi-channel review to the full MHD formulation that follows.


\section{PDEL formulation of thermoviscous resistive MHD}\label{sec:pdel}
\subsection{Unified state--history functional}
Let $\Omega(t)$ be a material domain generated by an orientation-preserving motion $\xx=\xx(\XX,t)$. All quantities below are consistently nondimensionalized, with magnetic permeability $\mu_0=1$. The independent thermal state consists of temperature $T>0$ and entropy density $\st$ per current volume. A representative PDEL functional is
\begin{equation}
\mathscr A=\int_{t_0}^{t_1}\!\int_{\Omega(t)}
\left[\frac12\rho|\vv|^2-\Phi(\rho,T)-T\st
-\frac12|\BB|^2-H_q+H_{\rm J}+H_\nu\right]\dd V\dd t.
\label{eq:action}
\end{equation}
The instantaneous terms represent kinetic, thermal and magnetic storage. This functional extends the reviewed state--history construction \citep{RenEntropy,RenSpending,RenEulerian,RenMultiChannel} to a retained induction field. The histories $H_q$, $H_{\rm J}$ and $H_\nu$ represent heat transport, resistive conversion and viscous conversion, respectively. Their meaning is fixed by the transport and differential rules below; their committed interior histories are not subjected to an ordinary variation.

Mass conservation is kinematic admissibility in the material-to-Eulerian description \citep{Newcomb1962,Holm1998,RenEulerian}:
\begin{equation}
\rho(\xx(\XX,t),t)\mathcal J(\XX,t)=\rho_{\rm ref}(\XX),\qquad
\mathcal J=\det\frac{\partial\xx}{\partial\XX}>0,
\qquad \vv\circ\xx=\partial_t\xx.
\label{eq:kinematics}
\end{equation}
Consequently $\partial_t\rho+\diver(\rho\vv)=0$. The material and conservative rates introduced in \eqref{eq:reviewrates} are, respectively,
\begin{equation}
\Dt=\partial_t+\vv\cdot\grad,\qquad
\Cv f=\partial_tf+\diver(f\vv).
\label{eq:rates}
\end{equation}
A history $H_D$ of power density $D$ is the spatial representation of a material accumulation:
\begin{equation}
H_D=\rho\Gamma_D,\qquad
\Dt\Gamma_D=\frac D\rho,\qquad
\Cv H_D=D.
\label{eq:historytransport}
\end{equation}
Equivalently, $\mathcal JH_D=\int_{t_0}^{t}\mathcal J(\XX,\tau)D(\xx(\XX,\tau),\tau)\dd\tau$ for zero initial history. A fixed-position time integral does not generally satisfy this identity.

For the ideal gas, using the mixed thermal state of PDEL \citep{RenEntropy,RenEulerian}, choose
\begin{equation}
\Phi(\rho,T)=\rho c_vT[1-\log(T/T_*)]
+\rho RT\log(\rho/\rho_*),\qquad
c_v=\frac{R}{\gamma-1},\quad \gamma>1.
\label{eq:freeenergy}
\end{equation}
Independent temperature variation subsequently yields $\st=-\Phi_T$, and hence
\begin{equation}
u=\Phi+T\st=\rho c_vT,\qquad
p=\rho\Phi_\rho-\Phi=\rho RT.
\label{eq:eos}
\end{equation}
The conjugacy is not substituted into \eqref{eq:action} before taking its mixed thermal differential. The specific entropy is $s=\st/\rho=c_v\log(T/T_*)-R\log(\rho/\rho_*)$, up to an arbitrary reference constant.

\subsection{Independent weak directions and common dissipative powers}
In a temporal-gauge chart, use $\BB=\curl\Avec$ and $\EE=-\partial_t\Avec$. The arbitrary vector-potential test $\atest=\delta\Avec$ is independent of the arbitrary fluid displacement $\xii$. The mechanical directions are the constrained Eulerian relabeling variations \citep{Holm1998,RenEulerian}; the magnetic test is kept independent, as required for a retained dynamical field \citep{Carlier,Gawlik2025}:
\begin{equation}
\begin{aligned}
\delta_\xi\vv&=\partial_t\xii+[\vv,\xii],&
\delta_\xi\rho&=-\diver(\rho\xii),\\
\delta_\xi\st&=-\diver(\st\xii),&
\delta_\xi T&=-\xii\cdot\grad T,
\end{aligned}
\label{eq:variations}
\end{equation}
where $[\vv,\xii]=(\vv\cdot\grad)\xii-(\xii\cdot\grad)\vv$. The displacement vanishes at the time endpoints. Compactly supported spatial tests establish the bulk equations; boundary data are specified in Section~\ref{sec:boundary}.

For a symmetric positive-definite resistivity tensor $\etab$, define
\begin{equation}
\boldsymbol r_A=\partial_t\Avec-\vv\times\BB,\qquad
\boldsymbol R_A=\etab^{-1}\boldsymbol r_A,\qquad
D_{\rm J}=\boldsymbol r_A\cdot\etab^{-1}\boldsymbol r_A.
\label{eq:relative}
\end{equation}
The associated weak power is
\begin{equation}
\mathcal W_{\rm J}[\xii,\atest]
=\int_{\Omega(t)}\boldsymbol R_A\cdot(\atest-\xii\times\BB)\dd V.
\label{eq:weakohm}
\end{equation}
When evaluated on the physical directions $\xii=\vv\delta t_m$ and $\atest=(\partial_t\Avec)\delta t_m$, the integrand becomes $D_{\rm J}\delta t_m$. This calibration does not restrict either arbitrary weak test.

For bulk viscosity and Fourier conduction, let
\begin{equation}
\ttau=\zeta(\diver\vv)\Id,\qquad
D_\nu=\ttau:\grad\vv=\zeta(\diver\vv)^2,\qquad
\qv=-\boldsymbol K\grad T,
\label{eq:viscousheat}
\end{equation}
with $\zeta>0$ and $\boldsymbol K=\boldsymbol K^T\succeq0$. The same stress is used in the weak mechanical pairing $\ttau:\grad\xii$ and in $D_\nu$. A nonnegative shear viscosity may also be retained, since its deviatoric strain rate vanishes on the exact solutions constructed below.

Each history in \eqref{eq:action} is defined by the push-forward of its constitutive integrand. With all integrands below evaluated at $(\xx(\XX,\tau),\tau)$,
\begin{equation}
\begin{aligned}
H_q(\xx(\XX,t),t)
&=\frac1{\mathcal J(\XX,t)}\int_{t_0}^{t}\mathcal J(\XX,\tau)\,
\diver(-\boldsymbol K\grad T)\dd\tau,\\
H_{\rm J}(\xx(\XX,t),t)
&=\frac1{\mathcal J(\XX,t)}\int_{t_0}^{t}\mathcal J(\XX,\tau)\,
\boldsymbol r_A\cdot\etab^{-1}\boldsymbol r_A\,\dd\tau,\\
H_\nu(\xx(\XX,t),t)
&=\frac1{\mathcal J(\XX,t)}\int_{t_0}^{t}\mathcal J(\XX,\tau)\,
\zeta(\diver\vv)^2\dd\tau.
\end{aligned}
\label{eq:explicitMHDhistories}
\end{equation}
After the vector-potential equation has established $\boldsymbol R_A=-\JJ$, the Joule-history integrand is equivalently $\JJ\cdot\etab\JJ$. This equivalence is a consequence of stationarity; the pre-stationarity definition in \eqref{eq:explicitMHDhistories} does not presuppose Ohm's law.

The PDEL history rules apply the common-power weak lifting reviewed in Section~\ref{sec:review} \citep{RenSpending,RenEulerian,RenMultiChannel}:
\begin{equation}
\begin{aligned}
\delta^{\rm ch}H_{\rm J}
&=D_{\rm J}\delta t_s-\boldsymbol R_A\cdot(\atest-\xii\times\BB),\\
\delta^{\rm ch}H_\nu&=D_\nu\delta t_s-\ttau:\grad\xii,\\
\delta^{\rm ch}H_q&=(\diver\qv)\delta t_s,
\end{aligned}
\label{eq:spending}
\end{equation}
with $\Cv H_q=\diver\qv$, $\Cv H_{\rm J}=D_{\rm J}$ and $\Cv H_\nu=D_\nu$. The entropy direction is $\delta^{\rm ch}\st=\Cv\st\,\delta t_s$, while the additional temperature variation $\delta T$ is arbitrary. Equations~\eqref{eq:variations} and \eqref{eq:spending} define distinct components of the mixed differential. In particular, the constitutive operator is evaluated at the current state; no additional variation of its past values is performed inside a locked history.

Ordinary variation of current storage, followed by the stated channel rules, gives the bulk expression
\begin{equation}
\begin{aligned}
\mathfrak D\mathscr A=\int\!\int_{\Omega(t)}\big\{&
[-\rho\Dt\vv-\grad p+\diver\ttau+\BB\times\boldsymbol R_A]\cdot\xii\\
&-(\JJ+\boldsymbol R_A)\cdot\atest-(\Phi_T+\st)\delta T\\
&+[-T\Cv\st-\diver\qv+D_{\rm J}+D_\nu]\delta t_s\big\}\dd V\dd t.
\end{aligned}
\label{eq:jointdifferential}
\end{equation}
Here $\JJ=\curl\BB$. The kinetic variation uses \eqref{eq:kinematics}; without mass admissibility its mechanical coefficient would additionally contain $-\vv[\partial_t\rho+\diver(\rho\vv)]$. The mixed thermal state contributes $-\grad p$ before elimination of entropy conjugacy.

Arbitrariness of the vector-potential test gives $\boldsymbol R_A=-\JJ$, so
\begin{equation}
\EE'=\EE+\vv\times\BB=\etab\JJ,
\qquad D_{\rm J}=\JJ\cdot\etab\JJ>0\quad(\JJ\ne0).
\label{eq:ohm}
\end{equation}
The mechanical and thermal tests give momentum balance and the entropy equation. Together with Faraday's identity and the equation of state, the resulting system is
\begin{subequations}\label{eq:system}
\begin{align}
\partial_t\rho+\diver(\rho\vv)&=0,\label{eq:mass}\\
\rho\Dt\vv&=-\grad p+\diver\ttau+\JJ\times\BB,\label{eq:mom}\\
\partial_t\BB&=\curl(\vv\times\BB)-\curl(\etab\JJ),
\quad\diver\BB=0,\label{eq:induction}\\
\rho c_v\Dt T+p\diver\vv&=D_{\rm J}+D_\nu-\diver\qv.
\label{eq:heat}
\end{align}
\end{subequations}
The independent vector-potential variation yields the vector Ohm equation. Magnetic storage is retained in \eqref{eq:action}, so Lorentz work and Joule heat arise from the same constitutive action without additional prescribed sources. The formulation assumes strictly positive resistivity and a compatible potential chart; the ideal-resistivity limit requires a different constrained admissible class. Its single-temperature Eulerian identities and fixed-box formulation are included in the Layer~3 formalization described in Section~\ref{sec:formal}. Related independent-potential dissipative formulations are developed in \citet{Carlier}; the present construction uses the PDEL weak-power and history pairing.

\subsection{Material textures and constitutive specification}
The anisotropic tensor is represented by a symmetric material field $\MM$, and $\beta$ is a scalar viscosity parameter. One objective transport prescription is
\begin{equation}
\Dt\MM=\boldsymbol W\MM-\MM\boldsymbol W,\qquad
\Dt\beta=0,\qquad \boldsymbol W=\skw\grad\vv.
\label{eq:textures}
\end{equation}
These fields carry no stored energy in the model and are not independent sources of mechanical or thermal power. They specify a heterogeneous conducting continuum. On a homogeneous expansion, $\boldsymbol W=0$, and they remain constant along each material trajectory.

The one-temperature construction will use
\begin{equation}
\etab=\eta_*(T/T_*)^{-2/3}\MM,\qquad
\zeta=\beta\rho(T/T_*)^{1/3},\qquad \eta_*>0.
\label{eq:oneclosures}
\end{equation}
Their material textures are determined in Sections~\ref{sec:realization} and \ref{sec:oneT} and are subsequently fixed. The exponent $-2/3$ is selected by the coupled construction, whereas the two-temperature extension in Section~\ref{sec:twoT} uses a different thermal state to obtain a $-3/2$ electron-temperature prefactor.


\section{Toroidal seed and passive resistive realization}\label{sec:realization}
\subsection{Reference equilibrium and magnetic shear}
In reference Cartesian coordinates $\XX=(X,Y,Z)$, consider the sheared equilibrium of \citet{Landreman} with the constant complex-phase extension of \citet{Issan}. For $\epsilon,S,\lambda>0$ and $\tau_s\in\mathbb R$, write
\begin{equation}
\begin{aligned}
w&=X+iY,&K&=\bar w\sqrt{1+\epsilon/\bar w^2},\\
\Xi&=wK+\frac\pi2-S+i\tau_s,&W_0&=\frac{i\sin\Xi}{2K}.
\end{aligned}
\label{eq:complexseed}
\end{equation}
The square-root branch is fixed by $\Re(K/\bar w)>0$ on the retained domain. The magnetic field and seed pressure are
\begin{equation}
\begin{aligned}
B_{0x}+iB_{0y}&=e^{-i\lambda Z}W_0,\\
B_{0z}&=\lambda^{-1}\Re(e^{-i\lambda Z}\cos\Xi),\\
\psi&=\tfrac12\{\sin^2(\lambda Z)+\lambda^2B_{0z}^2\},
\qquad p_0=p_a-\psi/\lambda^2.
\end{aligned}
\label{eq:seedfield}
\end{equation}
They satisfy the magnetostatic identities
\begin{equation}
\grad_X\!\cdot\BB_0=0,\qquad
\JJ_0=\grad_X\!\times\BB_0,\qquad
\JJ_0\times\BB_0=\grad_Xp_0.
\label{eq:mhs}
\end{equation}
The corresponding current is
\begin{equation}
\begin{aligned}
J_{0x}&=\left[\lambda+\frac{4Y^2-2\epsilon}{\lambda}\right]B_{0x}
-\frac{4XY}{\lambda}B_{0y},\\
J_{0y}&=-\frac{4XY}{\lambda}B_{0x}
+\left[\lambda+\frac{4X^2+2\epsilon}{\lambda}\right]B_{0y},\\
J_{0z}&=\lambda B_{0z}.
\end{aligned}
\label{eq:current}
\end{equation}
In particular, a smooth global design potential on this domain is
\begin{equation}
\Avec_0=\lambda^{-1}(B_{0x},B_{0y},0),\qquad
\grad_X\!\times\Avec_0=\BB_0,
\qquad g=\Avec_0\cdot\JJ_0.
\label{eq:designpotential}
\end{equation}
The potential identity follows directly from the $Z$-derivatives of \eqref{eq:seedfield} and the last line of \eqref{eq:current}.

For $m=2\psi$, the conditions
\begin{equation}
0<m_b<1,\qquad S>\arcsin\sqrt{m_b}
\label{eq:regularity}
\end{equation}
define a smooth embedded solid torus $\Omega_0$ bounded by $m=m_b$. Appendix~\ref{app:geometry} gives its explicit surface map, inverse-coordinate argument and volume element. All derivatives in the governing equations are Cartesian Euclidean derivatives. The field is nonzero on the closed domain, including the magnetic axis, and $\BB_0\cdot\nn_0=0$ on its boundary.

The rotational transform is evaluated from the field direction rather than prescribed as an independent profile, following the field-line construction of \citet{Landreman,Issan} and the magnetic-surface convention discussed in \citet{Helander2014}. With the period integrals $\mathcal P(m)$ and $\mathcal R(m)$ defined in Appendix~\ref{app:geometry},
\begin{equation}
\iota(m)=\frac{\mathcal R(m)}{\mathcal P(m)},\qquad
\Phi_{t,0}(m)=\frac1{4\lambda}\int_0^m\mathcal P(s)\dd s.
\label{eq:transform}
\end{equation}
Writing $q_s=1/\iota$ and $\rho_\Phi=\sqrt{\Phi_{t,0}(m)/\Phi_{t,0}(m_b)}$, the flux-normalized shear is
\begin{equation}
\widehat s=\frac{\dd\log q_s}{\dd\log\rho_\Phi}
=-2\frac{\int_0^m\mathcal P(s)\dd s}{\mathcal P(m)}\frac{\iota_m}{\iota}.
\label{eq:shear}
\end{equation}
For $\epsilon=\tau_s=1/2$ and $S=3$, Appendix~\ref{app:geometry} establishes $\iota_\psi(0)/\iota(0)>31/288$. Nonzero shear near the axis therefore follows analytically, independently of numerical differentiation of the plotted profile.

Table~\ref{tab:parameters} specifies the reference geometry and the two thermal examples before their first graphical use. The parameter sweeps in Section~\ref{sec:verification} vary one declared parameter at a time; all quantities are dimensionless.

\begin{table}[H]
\centering
\caption{Parameters of the one- and two-temperature examples.}
\label{tab:parameters}
\begin{tabular}{@{}lll@{}}\toprule
Quantity & One temperature & Two temperatures\\\midrule
$(\epsilon,S,\tau_s,\lambda)$ & $(1/2,3,1/2,4)$ & Same\\
Outer surface $m_b$ & $0.9$ & $0.9$\\
Seed-pressure offset $p_a$ & $1$ & $2$\\
$(b_*,T_*,R,\eta_*)$ & $(1,1,1,0.005)$ & Same\\
$(\gamma,c_v,\kappa_0,U)$ & $(5/3,3/2,0.08,0.32)$ & Same\\
Electron fraction $r$ & Not applicable & $1/4$\\
Analytical interval & $a\ge1$ (finite time) & $1\le a<4^{6/5}$\\
Displayed thermal interval & $1\le a\le4$ & $1\le a\le4$\\\bottomrule
\end{tabular}
\end{table}

\begin{figure}[tbp]
\centering
\includegraphics[width=.67\linewidth]{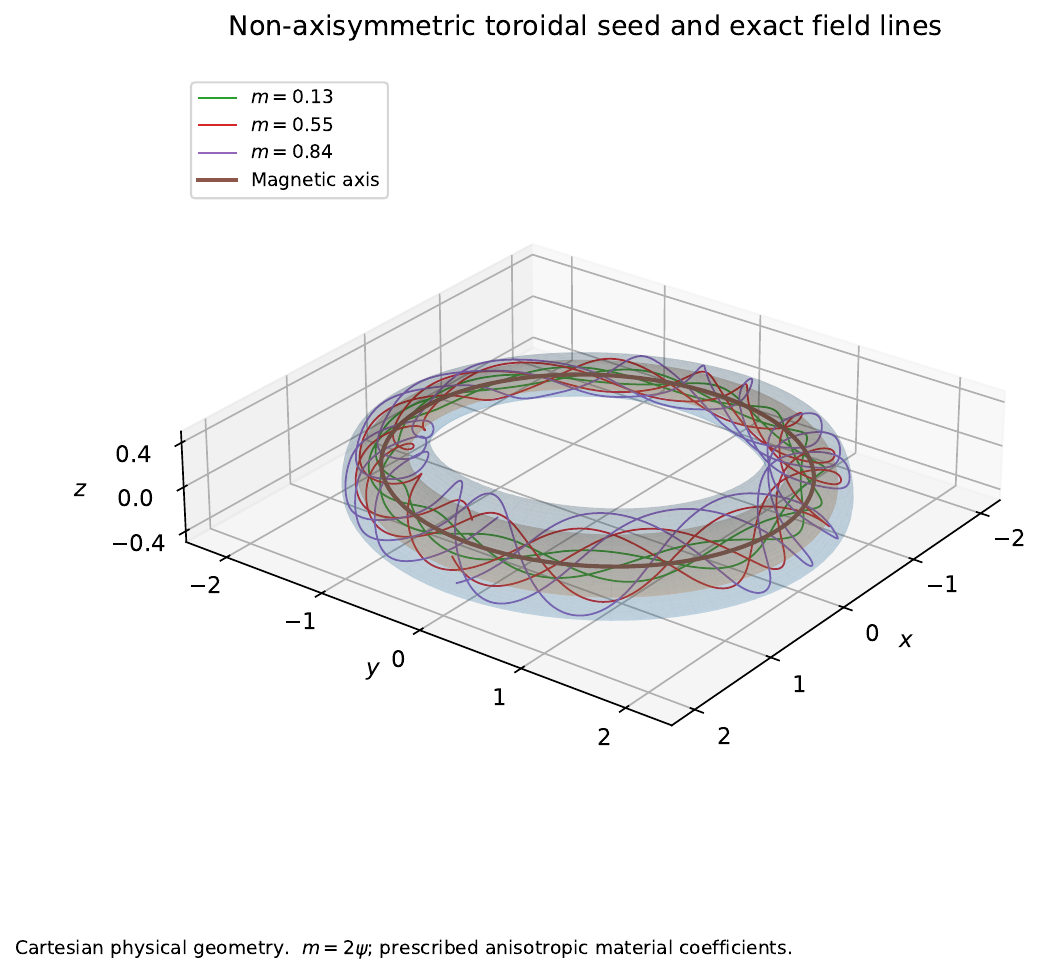}
\caption{Reference non-axisymmetric magnetic torus and field lines for the parameters in Table~\ref{tab:parameters}. The geometry is the attributed sheared Landreman--Issan equilibrium. The displayed field lines are evaluated by monotone inverse quadrature on the actual Euclidean magnetic surfaces.}
\label{fig:geometry}
\end{figure}

\subsection{A positive-definite tensor realizing the resistive field}
The next observation supplies the constitutive link between the reference geometry and its resistive evolution.
\begin{lemma}[Passive tensor realization]\label{lem:tensor}
Let $\BB_0=\grad_X\!\times\Avec_0$, $\JJ_0=\grad_X\!\times\BB_0$ and $g=\Avec_0\cdot\JJ_0>0$. For $\ell>0$, define
\begin{equation}
\MM_0=\ell\frac{\Avec_0\otimes\Avec_0}{g}
+\Id-\frac{\JJ_0\otimes\JJ_0}{|\JJ_0|^2}.
\label{eq:tensor}
\end{equation}
Then $\MM_0$ is symmetric positive definite and satisfies
\begin{equation}
\MM_0\JJ_0=\ell\Avec_0,\qquad
\grad_X\!\times(\MM_0\JJ_0)=\ell\BB_0.
\label{eq:eigenidentity}
\end{equation}
If the fields are smooth on a compact domain and $g$ is bounded away from zero, $\MM_0$ is uniformly positive definite there.
\end{lemma}
\begin{proof}
For every vector $\boldsymbol z$,
\[
\boldsymbol z^T\MM_0\boldsymbol z
=\ell\frac{(\boldsymbol z\cdot\Avec_0)^2}{g}
+|\boldsymbol P_{\JJ_0^\perp}\boldsymbol z|^2.
\]
Both terms are nonnegative. Their simultaneous vanishing requires $\boldsymbol z$ to be parallel to $\JJ_0$ and perpendicular to $\Avec_0$, which implies $\boldsymbol z=0$ because $g>0$. Multiplication of \eqref{eq:tensor} by $\JJ_0$ proves the first relation in \eqref{eq:eigenidentity}; taking its curl proves the second. Uniform positivity follows by continuity and compactness.
\end{proof}

For the present seed, choose $\ell=\lambda^2$. Equation~\eqref{eq:current} implies
\begin{equation}
g\ge\left(1-\frac{2\epsilon}{\lambda^2}\right)|\BB_{0\perp}|^2>0
\qquad\text{when}\qquad\lambda^2>2\epsilon,
\label{eq:gpositive}
\end{equation}
where $\BB_{0\perp}=(B_{0x},B_{0y},0)$. The horizontal current matrix is the sum of $\lambda\Id_2$, a positive-semidefinite rank-one matrix and $\lambda^{-1}\operatorname{diag}(-2\epsilon,2\epsilon)$, which proves the bound. The explicit surface map gives $|\BB_{0\perp}|>0$ throughout the regular torus.

The tensor is designed once from the reference potential and then treated as fixed material data. Changing the design potential by a gauge gradient generally defines a different material tensor; it is not a transformation of the same constitutive model. Ordinary electromagnetic gauge freedom remains available for the evolving potential. Likewise, \eqref{eq:eigenidentity} is a differential identity with the compatible boundary traces used below, rather than a spectral assertion for a homogeneous perfectly conducting boundary problem.

Positive definiteness establishes passivity, not microscopic realizability. For example, conventional symmetric magnetic-field-aligned transport distinguishes parallel and perpendicular components \citep{Braginskii1965,Poulipoulis}, giving a tensor of the form
\begin{equation}
\etab=\eta_\parallel\widehat{\BB}\otimes\widehat{\BB}
+\eta_\perp(\Id-\widehat{\BB}\otimes\widehat{\BB}),
\label{eq:gyrotropic}
\end{equation}
and consequently maps $\JJ$ into $\operatorname{span}\{\JJ,\BB\}$. The present selected resistive field requires $\MM_0\JJ_0=\lambda^2\Avec_0$, but $\Avec_0\cdot(\JJ_0\times\BB_0)$ is not identically zero. At the example point with labels $(m,\theta,\zeta)=(0.4,0.43,0.71)$ it is approximately $-0.00565946$. Thus \eqref{eq:gyrotropic} cannot in general produce this particular relative electric field. An alternative electric-potential gradient or a different magnetic trajectory may alter that conclusion; it is not a nonexistence theorem for all field-aligned anisotropic models.

\subsection{Restriction on a scalar-resistivity realization}
The tensorial freedom in \eqref{eq:tensor} is essential to this particular shape-preserving construction.
\begin{proposition}[Seed-specific scalar obstruction]\label{prop:scalar}
For the seed \eqref{eq:seedfield} with \eqref{eq:regularity} and $\lambda^2>2\epsilon$, no positive scalar $r(\XX)$ satisfies
\begin{equation}
\grad_X\!\times(r\JJ_0)=\Lambda\BB_0
\label{eq:scalareigen}
\end{equation}
in a neighborhood of its outer magnetic-axis point for a spatial constant $\Lambda$.
\end{proposition}
\begin{proof}
Dotting \eqref{eq:scalareigen} with $\JJ_0$ fixes the only possible ratio $r/\Lambda$ wherever $\JJ_0\cdot(\grad_X\!\times\JJ_0)\ne0$. At the axis point $(x_a,0,0)$, both this denominator and $\JJ_0\cdot\BB_0$ are positive. Appendix~\ref{app:scalar} evaluates the full curl of the resulting ratio and obtains the strictly positive residual component
\begin{equation}
\left[\grad_X\!\times\left(\frac r\Lambda\JJ_0\right)-\BB_0\right]_z
=\frac{16x_a^3\cosh\tau_s}
{\lambda k_a(\lambda^2+4x_a^2)^2}>0,
\label{eq:scalarresidual}
\end{equation}
where $k_a=\sqrt{x_a^2+\epsilon}$ and $x_ak_a=S$. This contradicts \eqref{eq:scalareigen}.
\end{proof}
A homologous evolution that preserves this seed shape would require \eqref{eq:scalareigen} at each time. The proposition therefore excludes that scalar-resistive path even when the scalar coefficient is spatially nonuniform. It does not exclude other toroidal seeds, changing magnetic shapes or nonhomologous velocities.


\section{One-temperature exact evolution}\label{sec:oneT}
\subsection{Construction and complete field equations}
The exponent relations, stress and power identities, and material-history derivatives are accompanied by symbolic calculations and independent Cartesian residual tests in the supplementary files. The scope of the Lean formalization is given in Section~\ref{sec:formal}.

\begin{theorem}[Resistive thermoviscous homologous evolution]\label{thm:oneT}
Let the reference seed satisfy \eqref{eq:mhs}, \eqref{eq:designpotential}, $g>0$ and $p_0>g/3$ on $\overline\Omega_0$. Choose $\gamma=5/3$, $b_*,T_*,\eta_*>0$ and set
\begin{equation}
\kappa_0=\lambda^2\eta_*,\qquad U=4\kappa_0,\qquad
a(t)=1+Ut,\qquad \XX=\xx/a(t).
\label{eq:scale}
\end{equation}
Define the reference density and viscosity texture by
\begin{equation}
\rho_0=\frac{b_*^2(12p_0-g)}{6c_vT_*},\qquad
\beta_0=\frac{b_*^2(3p_0-g)}{108\kappa_0\rho_0}.
\label{eq:materialdata}
\end{equation}
On $\Omega(t)=a(t)\Omega_0$, the fields
\begin{equation}
\begin{aligned}
\vv&=U\XX,&\rho&=a^{-3}\rho_0(\XX),\\
\BB&=b_*a^{-9/4}\BB_0(\XX),&\JJ&=b_*a^{-13/4}\JJ_0(\XX),\\
T&=T_*a^{-3/2},&p&=\rho RT,\\
\MM&=\MM_0(\XX),&\beta&=\beta_0(\XX)
\end{aligned}
\label{eq:onesolution}
\end{equation}
solve \eqref{eq:system} and \eqref{eq:textures} with \eqref{eq:oneclosures} for every finite $t\ge0$. Density, temperature, thermodynamic pressure, resistivity and bulk viscosity are positive, and both resistive and viscous powers are strictly positive.
\end{theorem}
\begin{proof}
The motion $\xx=a\XX$ has Jacobian $a^3$, so \eqref{eq:mass} follows directly. Its material acceleration is $\Dt\vv=\ddot a\XX=0$. The Eulerian time derivative and convective acceleration are individually nonzero and cancel. The velocity gradient is $(U/a)\Id$, with zero spin and zero deviatoric part; hence \eqref{eq:textures} is satisfied.

The viscosity and isotropic viscous stress $\ttau=\Pi\Id$ are
\begin{equation}
\zeta=\frac{b_*^2(3p_0-g)}{108\kappa_0}a^{-7/2},\qquad
\Pi=\frac{b_*^2(3p_0-g)}9a^{-9/2}.
\label{eq:viscoussolution}
\end{equation}
Since $R=2c_v/3$, the thermodynamic pressure obeys
\begin{equation}
p-\Pi=b_*^2a^{-9/2}p_0,\qquad
\JJ\times\BB=b_*^2a^{-11/2}\grad_Xp_0.
\label{eq:forcebalance}
\end{equation}
The three momentum components therefore vanish exactly.

From Lemma~\ref{lem:tensor}, the relative electric field is
\begin{equation}
\EE'=\etab\JJ=\kappa_0b_*a^{-9/4}\Avec_0(\XX),
\qquad
\curl\EE'=\frac{\kappa_0}{a}\BB.
\label{eq:relativefield}
\end{equation}
For homogeneous expansion the induction equation becomes
\[
\Dt\BB+\frac{2U}{a}\BB=-\curl\EE'.
\]
The left side evaluated from \eqref{eq:onesolution} equals $-U\BB/(4a)$, which is the right side because $U=4\kappa_0$. The solenoidal constraint follows from that of the seed.

The realized powers are
\begin{equation}
D_{\rm J}=\kappa_0b_*^2g\,a^{-11/2},\qquad
D_\nu=\frac43\kappa_0b_*^2(3p_0-g)a^{-11/2}.
\label{eq:powers}
\end{equation}
On the thermal side,
\begin{equation}
\rho c_v\Dt T+p\diver\vv
=\frac{\kappa_0b_*^2}{3}(12p_0-g)a^{-11/2}
=D_{\rm J}+D_\nu.
\label{eq:thermalclosure}
\end{equation}
Temperature is spatially uniform, so the Fourier flux vanishes for any retained positive-semidefinite conductivity tensor. The hypotheses give positivity of \eqref{eq:materialdata}, \eqref{eq:viscoussolution} and \eqref{eq:powers}.
\end{proof}

An explicit potential pair useful for independent verification is
\begin{equation}
\Avec=b_*a^{-5/4}\Avec_0(\XX),\qquad
\phi=\vv\cdot\Avec,\qquad
\EE=-\partial_t\Avec-\grad\phi.
\label{eq:potentialpair}
\end{equation}
It produces the same electric field as \eqref{eq:relativefield} and is gauge-equivalent to the temporal-gauge chart used in Section~\ref{sec:pdel}.

The evolving thermal pressure is
\begin{equation}
p=\frac{b_*^2}{9}(12p_0-g)a^{-9/2}.
\label{eq:thermalpressure}
\end{equation}
It is generally not constant on magnetic surfaces because $g$ is not a magnetic-surface function. The surface-aligned quantity is the effective pressure $p-\Pi$ in \eqref{eq:forcebalance}. Bulk viscosity is therefore required by the vector force balance as well as by the thermal equation; it cannot be removed while retaining the same fields.

\subsection{Constitutive exponent selection}
The exponents in Theorem~\ref{thm:oneT} follow from a restricted but explicit matching problem. Let $T=T_*a^{s_T}$, $\BB=b_*a^{\delta_B-2}\BB_0$, and $\etab=\eta_*(T/T_*)^\nu\MM_0$, with a linear scale factor and common pressure, viscous-stress and magnetic-stress exponents. Appendix~\ref{app:exponents} gives
\begin{equation}
s_T=\frac1\nu,\qquad
\delta_B=\frac{\nu+1}{2\nu},\qquad
U=-\frac{2\nu\kappa_0}{\nu+1},\qquad
q_\zeta=\nu+1,
\label{eq:exponents}
\end{equation}
where $\zeta=\beta_0\rho(T/T_*)^{q_\zeta}$ and $\nu\notin\{0,-1\}$. For $p_0,g>0$, a sufficiently large positive seed-pressure offset yields a passive expanding branch precisely in the large-offset window
\begin{equation}
\gamma>\frac43,\qquad -1<\nu<-\frac1{3(\gamma-1)}.
\label{eq:passivewindow}
\end{equation}
At $\gamma=5/3$, the choice $\nu=-2/3$ belongs to this interval and gives \eqref{eq:onesolution}. By contrast, substituting $\nu=-3/2$ into this same separation class produces contraction with a positive isotropic viscous stress, requiring a negative bulk-viscosity coefficient. The two-temperature construction below changes the thermal state space rather than changing the exponent in an otherwise unchanged one-temperature solution.

\subsection{Material flux loss and retained magnetic shear}
Under \eqref{eq:onesolution}, the direction of the field in material coordinates is unchanged. The transform and normalized shear on corresponding material surfaces are consequently retained, while magnetic flux through a material cut decreases:
\begin{equation}
\iota(m,t)=\iota(m,0),\qquad
\widehat s(m,t)=\widehat s(m,0),\qquad
\Phi_t(m,t)=b_*a^{-1/4}\Phi_{t,0}(m).
\label{eq:fluxshear}
\end{equation}
The area factor is $a^2$, whereas the field amplitude is $a^{-9/4}$. The additional $a^{-1/4}$ flux factor distinguishes the solution from ideal frozen-in expansion. Thus nonzero magnetic shear coexists with a nonzero resistive induction term, although the shear profile is not redistributed in time.

Field-line winding is preserved by the shape-preserving resistive operator, while the field amplitude decays faster than the ideal $a^{-2}$ dilution. The solution thus describes resistive flux loss without rearrangement of the magnetic surfaces. It makes no assertion about magnetic reconnection or confinement performance. Relative to the magnetostatic seeds of \citet{Landreman,Issan}, the evolution also determines current, stress, temperature and irreversible power on the moving torus.

\begin{figure}[tbp]
\centering
\includegraphics[width=.67\linewidth]{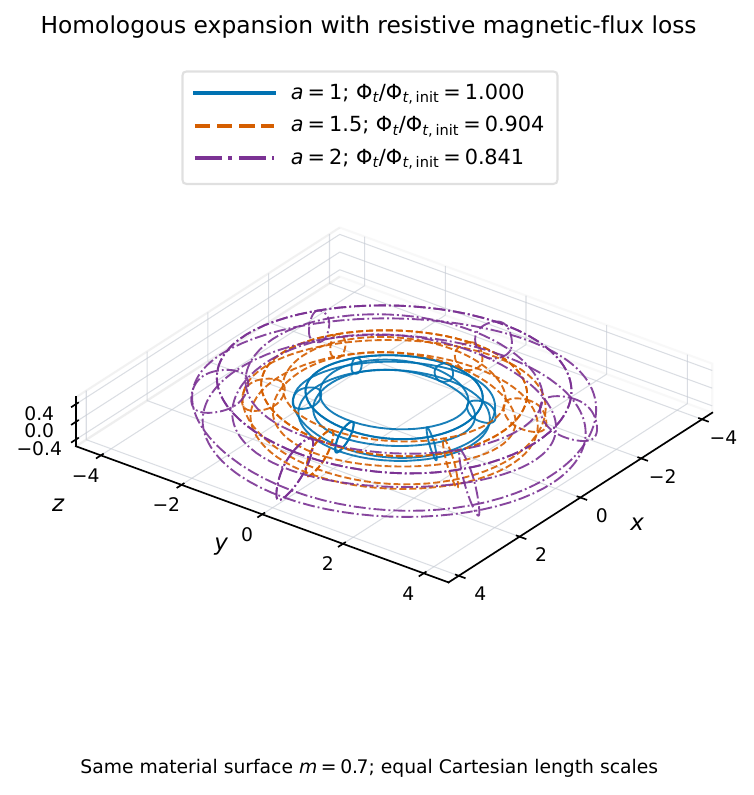}
\caption{Homologous expansion of the same material magnetic surface $m=0.7$. Blue solid, orange dashed and purple dash--dot meshes denote $a=1$, $1.5$ and $2$, respectively; the legend also gives their material-flux ratios $a^{-1/4}$. The mesh curves are surface-coordinate lines, not magnetic field-line trajectories. Equal Cartesian length scales are retained. The magnetic evolution is common to both thermal branches.}
\label{fig:expansion}
\end{figure}


\section{Two-temperature extension}\label{sec:twoT}
\subsection{Thermal channels and a positive exchange law}
A two-temperature state permits the same magnetic and mechanical evolution to use an electron-temperature resistivity prefactor $T_e^{-3/2}$. Let
\begin{equation}
c_e=c_i=\frac{c_v}{2},\qquad R_e=R_i=\frac R2,\qquad
\overline T=\frac{T_e+T_i}{2},\qquad
p=p_e+p_i=\rho R\overline T,
\label{eq:twostate}
\end{equation}
where $p_j=\rho R_jT_j$. The coefficients are per unit mixture mass. Both thermal components share the bulk velocity; separate electron and ion momenta or Hall drift are not part of this model.

Replace the thermal block in \eqref{eq:action} by
\[
\Phi_e(\rho,T_e)+T_e\st_e+\Phi_i(\rho,T_i)+T_i\st_i,
\]
with each free energy of the form \eqref{eq:freeenergy} using $c_j,R_j$. Independent thermal variations yield $\st_j=-\partial_{T_j}\Phi_j$. Assign Joule heat to electrons and viscous heat to ions, and use one exchange power with opposite signs in the two channels:
\begin{equation}
\begin{aligned}
\rho c_e\Dt T_e+p_e\diver\vv&=D_{\rm J}+Q_{i\to e},\\
\rho c_i\Dt T_i+p_i\diver\vv&=D_\nu-Q_{i\to e},\\
Q_{i\to e}&=G_{ie}(T_e,\XX)(T_i-T_e).
\end{aligned}
\label{eq:twoheat}
\end{equation}
Heat conduction can be retained for either species, but vanishes on the spatially uniform temperatures constructed here.

Choose $0<r<1$, put $T_{e0}=rT_*$ and define the material scalar
\begin{equation}
C_0(\XX)=\frac{4r}{9}(12p_0-g),\qquad
a_e=(T_e/T_{e0})^{-3/2}.
\label{eq:Czero}
\end{equation}
The constitutive laws are
\begin{equation}
\begin{aligned}
\etab&=\eta_*(T_e/T_{e0})^{-3/2}\MM,\\
\zeta&=\beta\rho\left(\frac{T_e+T_i}{2T_*}\right)^{1/3},\\
G_{ie}(T_e,\XX)&=\frac{\kappa_0b_*^2}{2T_*}a_e^{-4}
\frac{C_0(\XX)a_e^{5/6}-g(\XX)}{1-r a_e^{5/6}}.
\end{aligned}
\label{eq:twoclosures}
\end{equation}
Here $g$ and $C_0$ are fixed advected scalar textures. The exchange law depends on current electron temperature and material data, rather than laboratory time or a separately imposed heat source. Its specified passive state band is
\begin{equation}
\left(\frac{g}{C_0}\right)^{6/5}<a_e<r^{-6/5},
\label{eq:exchangeband}
\end{equation}
on which $G_{ie}>0$. The distinction between a temperature scaling and a collision-derived transport law is important \citep{Spitzer,Braginskii1965,Lingam}. Only the temperature prefactor in \eqref{eq:twoclosures} has the Spitzer exponent \citep{Spitzer}; neither $\MM$ nor $G_{ie}$ is derived from the classical collisional closure.

\subsection{Exact solution and entropy production}
\begin{theorem}[Finite-interval two-temperature realization]\label{thm:twoT}
Retain the seed, density, velocity and magnetic fields of Theorem~\ref{thm:oneT}, and suppose $p_0>g/3$ and $C_0>g>0$. Replace the temperature by
\begin{equation}
T_e=rT_*a^{-2/3},\qquad
T_i=2T_*a^{-3/2}-rT_*a^{-2/3}.
\label{eq:twotemperatures}
\end{equation}
For $1\le a<r^{-6/5}$, these fields satisfy mass, momentum, induction and both thermal equations \eqref{eq:twoheat} with \eqref{eq:twoclosures}. The temperatures, resistivity tensor, bulk viscosity and exchange coefficient are positive, with $T_i>T_e$ and $Q_{i\to e}>0$.
\end{theorem}
\begin{proof}
The mean temperature is $\overline T=T_*a^{-3/2}$. Total pressure, internal energy, viscosity and momentum therefore coincide with the one-temperature solution. In addition,
\[
(T_e/T_{e0})^{-3/2}=a,
\]
so the resistive field, induction equation and Joule power are unchanged. Substitution into the electron equation requires
\begin{equation}
Q_{i\to e}=\kappa_0b_*^2a^{-11/2}(C_0a^{5/6}-g).
\label{eq:exchangepower}
\end{equation}
Since $T_i-T_e=2T_*a^{-3/2}(1-ra^{5/6})$, \eqref{eq:twoclosures} gives exactly this power. The ion equation follows by subtracting the electron equation from the already established total thermal equation. The inequalities follow from $C_0>g$ and $a<r^{-6/5}$.
\end{proof}

With specific entropies $s_j=c_j\log T_j-R_j\log\rho+\text{constant}$, the total entropy balance is
\begin{equation}
\rho\Dt(s_e+s_i)
=\frac{D_{\rm J}}{T_e}+\frac{D_\nu}{T_i}
+G_{ie}\frac{(T_i-T_e)^2}{T_eT_i}>0.
\label{eq:twoentropy}
\end{equation}
The exchange conserves total thermal energy, but contributes positively to entropy production. The exact species entropy changes are
\begin{equation}
\begin{aligned}
s_e-s_{e0}&=\frac{2c_v}{3}\log a,\\
s_i-s_{i0}&=\frac{c_v}{2}\log\frac{T_i}{(2-r)T_*}+c_v\log a.
\end{aligned}
\label{eq:speciesentropy}
\end{equation}
The ion entropy attains its maximum at $a=(3/4r)^{6/5}$ when this point lies after the initial state. It subsequently decreases when outgoing exchange exceeds viscous heat input, consistently with \eqref{eq:twoentropy}. Transfer to the colder electrons maintains nonnegative total entropy production. These relations provide coupled checks of the exchange signs and temperature weights. The exchange coefficient is inverse-designed for the trajectory and is not a Coulomb-collision rate.

\subsection{Nonremovable endpoint of the exchange law}
The upper endpoint in Theorem~\ref{thm:twoT} is a constitutive restriction, not only a numerical stiffness threshold. Put $a_{\rm eq}=r^{-6/5}$. As $a\to a_{\rm eq}^{-}$, both temperatures approach the same positive value, but \eqref{eq:exchangepower} tends to
\begin{equation}
Q_{\rm eq}=\kappa_0b_*^2a_{\rm eq}^{-11/2}N_{\rm eq},\qquad
N_{\rm eq}=\frac{C_0}{r}-g=\frac{48p_0-13g}{9}>\frac g3>0.
\label{eq:Qendpoint}
\end{equation}
Furthermore,
\begin{equation}
T_i-T_e=\frac{5T_*}{3}a_{\rm eq}^{-5/2}(a_{\rm eq}-a)
+O((a_{\rm eq}-a)^2).
\label{eq:Tendpoint}
\end{equation}
Consequently,
\begin{equation}
G_{ie}\sim\frac{3\kappa_0b_*^2}{5T_*}
\frac{a_{\rm eq}^{-3}N_{\rm eq}}{a_{\rm eq}-a}.
\label{eq:Gpole}
\end{equation}
The pole is unavoidable under the stated assumptions. A bounded exchange coefficient cannot maintain this temperature path through equality while supplying the nonzero required exchange power. Regularizing the coefficient would change the exact thermal equations and would require a different temperature trajectory or power partition.


\section{Energy balance and transported power histories}\label{sec:energy}
The flux evolution in \eqref{eq:fluxshear} is accompanied by internal conversion and boundary power transfer. The following identities express these processes in terms of the exact fields and their constitutive responses.
\subsection{Local energy and the moving boundary}
For the one-temperature branch let $u=\rho c_vT$, and for the two-temperature branch let $u=\rho(c_eT_e+c_iT_i)$. Define
\begin{equation}
\mathcal E=\frac12\rho|\vv|^2+u+\frac12|\BB|^2.
\label{eq:energydef}
\end{equation}
Using the magnetic/thermal power split familiar from resistive MHD \citep{Carlier,Gawlik2025}, taking the scalar product of momentum with $\vv$, contracting induction with $\BB$, and adding the thermal equation gives, on both exact branches,
\begin{equation}
\partial_t\mathcal E+
\diver\left[(\tfrac12\rho|\vv|^2+u+p)\vv
-\ttau\vv+\EE\times\BB\right]=0.
\label{eq:localenergy}
\end{equation}
The heat flux is zero on these solutions. For general nonuniform temperatures, its contribution would be added to the energy flux. Resistive and viscous powers cancel between their donor equations and the thermal receiver. In the two-temperature case, exchange also cancels on adding the thermal channels, without cancelling its entropy production.

Let the total Cauchy stress be
\begin{equation}
\boldsymbol{\mathcal T}=-p\Id+\ttau+\BB\otimes\BB-\frac12|\BB|^2\Id.
\label{eq:totalstress}
\end{equation}
Combining Reynolds transport with the electromagnetic bulk--boundary power pairing \citep{PortHamiltonian,RenMultiChannel}, on $\Omega(t)$ one obtains
\begin{equation}
\frac{\dd}{\dd t}\int_{\Omega(t)}\mathcal E\dd V
=-\int_{\partial\Omega(t)}
[-\boldsymbol{\mathcal T}\vv+\EE'\times\BB]\cdot\nn\dd S.
\label{eq:movingenergy}
\end{equation}
The relative electric field $\EE'=\EE+\vv\times\BB$ is required alongside total traction work. This distinguishes energy closure from constancy of energy in a materially moving volume.

Define the reference integrals
\begin{equation}
I_p=\int_{\Omega_0}p_0\dd V,\qquad
I_g=\int_{\Omega_0}g\dd V,\qquad
I_B=\int_{\Omega_0}|\BB_0|^2\dd V.
\label{eq:integrals}
\end{equation}
Two reference identities make the boundary power explicit:
\begin{equation}
\begin{aligned}
\int_{\partial\Omega_0}(p_0+|\BB_0|^2/2)\XX\cdot\nn_0\dd S
&=3I_p+I_B/2,\\
\int_{\partial\Omega_0}(\Avec_0\times\BB_0)\cdot\nn_0\dd S
&=I_B-I_g.
\end{aligned}
\label{eq:referenceports}
\end{equation}
The first follows by testing the divergence-free seed Cauchy stress against $\XX$, using $\BB_0\cdot\nn_0=0$. The second follows from $\grad_X\cdot(\Avec_0\times\BB_0)=|\BB_0|^2-g$.

The outward mechanical and electromagnetic powers are therefore
\begin{equation}
\begin{aligned}
P_{\rm mech}^{\rm out}&=4\kappa_0b_*^2a^{-5/2}(3I_p+I_B/2),\\
P_{\rm em}^{\rm out}&=\kappa_0b_*^2a^{-5/2}(I_B-I_g).
\end{aligned}
\label{eq:outwardpowers}
\end{equation}
The total kinetic energy $K_0=\frac12\int_{\Omega_0}\rho_0U^2|\XX|^2\dd V$ is constant. Direct volume integration gives
\begin{equation}
E(t)=K_0+b_*^2a^{-3/2}
\left[\frac{12I_p-I_g}{6}+\frac{I_B}{2}\right],
\label{eq:integratedenergy}
\end{equation}
and hence
\begin{equation}
-\dot E=P_{\rm mech}^{\rm out}+P_{\rm em}^{\rm out}
=\kappa_0b_*^2a^{-5/2}(12I_p-I_g+3I_B).
\label{eq:globalenergy}
\end{equation}
The sign of the electromagnetic port is determined by $I_B-I_g$; it is not assumed to be outward. Consequently, decreasing magnetic energy cannot in general be identified with Joule heating alone: moving-boundary mechanical work and relative Poynting transfer also enter the budget. For the baseline geometry the electromagnetic contribution is inward, yet mechanical work is larger and the material-domain total energy decreases. The total-energy balance therefore describes an open material system.

\subsection{Compatible initial and boundary data}\label{sec:boundary}
At $t=0$, prescribe the selected exact fields and the material textures. On the material wall, prescribe the velocity $\vv=U\XX$ and the tangential relative electric trace furnished by \eqref{eq:relativefield}. Its surface-curl compatibility maintains $\BB\cdot\nn=0$. The heat-flux data are homogeneous. The corresponding total traction is an output, not an additional independent condition imposed together with the velocity.

These boundary data support \eqref{eq:movingenergy}, but the wall is generally not a perfect conductor because its tangential relative electric field is nonzero. The reference surface is material and impermeable to mass, not energetically isolated. No exterior vacuum, coil circuit or material-wall response is included. The construction provides compatible exact initial and boundary data rather than a well-posedness theorem for arbitrary nearby data.

\subsection{Reference-volume histories and entropy}
Let a hat indicate power accumulated per reference volume. Direct integration of \eqref{eq:powers} with the material Jacobian $a^3$ gives
\begin{equation}
\widehat H_{\rm J}=\frac{b_*^2g}{6}(1-a^{-3/2}),\qquad
\widehat H_\nu=\frac{2b_*^2}{9}(3p_0-g)(1-a^{-3/2}).
\label{eq:histories}
\end{equation}
The current-volume densities are $H_j(\xx,t)=a^{-3}\widehat H_j(\XX,t)$ and satisfy $\Cv H_j=D_j$. Thus the accumulation and its dilatational dilution are distinguished as in the Eulerian PDEL history construction \citep{RenEulerian}, as required by \eqref{eq:historytransport}.

Writing $u_0=\rho_0c_vT_*=b_*^2(12p_0-g)/6$, the complete material internal-energy budget is
\begin{equation}
\begin{aligned}
\widehat H_{\rm J}+\widehat H_\nu&=\frac{u_0}{3}(1-a^{-3/2}),\\
\widehat W_p:=\int_0^t a^3p\diver\vv\dd\tau
&=\frac{4u_0}{3}(1-a^{-3/2}),\\
a^3u-u_0&=\widehat H_{\rm J}+\widehat H_\nu-\widehat W_p.
\end{aligned}
\label{eq:materialbudget}
\end{equation}
Equation~\eqref{eq:materialbudget} gives the material energy partition. With $\Delta=1-a^{-3/2}$, the material element receives irreversible heat $u_0\Delta/3$, performs pressure work $4u_0\Delta/3$, and therefore loses internal energy $u_0\Delta$. Thus pressure work is exactly four times the accumulated irreversible heat along this branch. At $a=2$, the internal energy per reference volume is $2^{-3/2}\simeq0.35355$ of its initial value even though both dissipative histories have increased. Cooling here measures net energy loss through expansion, not absence of irreversible conversion. The one-temperature specific entropy satisfies
\begin{equation}
s-s_0=\frac{c_v}{2}\log a,\qquad
\rho T\Dt s=D_{\rm J}+D_\nu>0.
\label{eq:oneentropy}
\end{equation}

For the two-temperature branch, the exchange history is
\begin{equation}
\widehat H_{ie}=b_*^2
\left[\frac{3C_0}{8}(1-a^{-2/3})-\frac g6(1-a^{-3/2})\right],
\qquad \partial_t\widehat H_{ie}=a^3Q_{i\to e}.
\label{eq:exchangehistory}
\end{equation}
It enters the electron receiver and ion donor with opposite signs. There is one physical exchange, not an additional total-energy source. Its entropy contribution is $Q_{i\to e}(T_e^{-1}-T_i^{-1})\ge0$.

Finally, the normalized local history $h=6\widehat H_{\rm J}/(b_*^2g)=1-a^{-3/2}$ reconstructs the one-temperature trajectory:
\begin{equation}
\begin{aligned}
a&=(1-h)^{-2/3},&T/T_*&=1-h,&\rho/\rho_0&=(1-h)^2,\\
\BB(a\XX,t)&=b_*(1-h)^{3/2}\BB_0(\XX),&&
\Phi_t/\Phi_{t,\mathrm{init}}&=(1-h)^{1/6}.
\end{aligned}
\label{eq:historycoordinate}
\end{equation}
This is an explicit state coordinate on the constructed solution manifold. Since it can be recovered from the instantaneous scale or temperature, it does not introduce an additional hereditary material law.

The distinction between reference and current histories also has an observable consequence. Along a material point, $\widehat H_{\rm J}$ increases monotonically, whereas its current-volume density is proportional to $a^{-3}(1-a^{-3/2})$ and reaches a maximum at $a=(3/2)^{2/3}$. Subsequent decrease of $H_{\rm J}$ reflects dilatational dilution of accumulated energy per current volume, not negative Joule power. This separates a transported PDEL history from a fixed-location temporal accumulation.

The resulting verification targets comprise effective pressure $p-\Pi$, magnetic-flux decay, cancellation of internal power transfers, material-history transport and boundary flux. Their coupled evaluation with a common set of constitutive data complements exact-equilibrium benchmarks and energy-consistent MHD discretizations \citep{Landreman,Issan,Gawlik2021,Gawlik2025}. These tests can identify inconsistent coupling and omitted transport terms; numerical accuracy and stability require separate analysis.


\section{Parameter-dependent results and verification}\label{sec:verification}
\subsection{Reference settings and admissible parameter families}
Table~\ref{tab:parameters}, introduced with the reference geometry, specifies the baseline examples. The two-temperature case changes the pressure offset and therefore the material density and viscosity, but retains the same magnetic surfaces. The parameters are nondimensional benchmark values, not a fit to a plasma experiment. Conservative bounds on the entire baseline reference domain are
\begin{equation}
1<G<5,\qquad X^2+Y^2<5,\qquad
|\BB_{0\perp}|^2<1,\qquad 0<g<\frac{37}{16}.
\label{eq:domainbounds}
\end{equation}
For $p_a=1$, $p_0>31/32$, hence
\begin{equation}
3p_0-g>\frac{19}{32},\qquad \rho_0>\frac{149}{144}>0.
\label{eq:passivitybounds}
\end{equation}
For $p_a=2$ and $r=1/4$, $p_0>63/32$ and
\begin{equation}
C_0-g=\frac{12p_0-10g}{9}>\frac1{18}>0.
\label{eq:twopassivity}
\end{equation}
These inequalities establish passivity independently of the sampled matrix eigenvalues. Table~\ref{tab:sweeps} lists the additional comparisons. Changing a parameter generally changes the inverse-designed material textures; the comparisons are between specified exact realizations, not perturbations of one fixed constitutive medium.
\begin{table}[H]
\centering\small
\caption{Parameter studies. Unlisted parameters retain Table~\ref{tab:parameters}; the phase study uses $p_a=2$ and the two-temperature studies use the two-temperature baseline.}
\label{tab:sweeps}
\begin{tabularx}{\linewidth}{@{}p{.17\linewidth}p{.32\linewidth}X@{}}\toprule
Parameter & Values or range & Isolated effect\\\midrule
Phase $\tau_s$ & $0,\ 0.25,\ 0.5,\ 0.75$ & Radial winding and shear; $\tau_s=0$ is the Landreman magnetic reference.\\
Resistive scale $\eta_*$ & $0.0025,\ 0.005,\ 0.01$ & Physical-time rate; $\beta_0$ and $U$ change consistently.\\
Pressure offset $p_a$ & $0.85\le p_a\le2.5$ & Density, bulk viscosity and heat partition; geometry fixed.\\
Electron fraction $r$ & $0.25,\ 0.375,\ 0.5$ & Temperature separation, ion entropy and interval endpoint.\\\bottomrule
\end{tabularx}
\end{table}
For each family, the code records the domain parameters and conservative positivity margins. In particular, the lower pressure offset still satisfies $3p_0-g>0$ using \eqref{eq:domainbounds}, and increasing $r$ above $1/4$ at $p_a=2$ preserves the initial exchange margin. Every two-temperature curve stops strictly below its own endpoint.

\subsection{Three-dimensional distributions of the exact fields}\label{sec:fields3d}
\begin{figure}[tbp]
\centering
\begin{subfigure}{.49\linewidth}
\includegraphics[width=\linewidth]{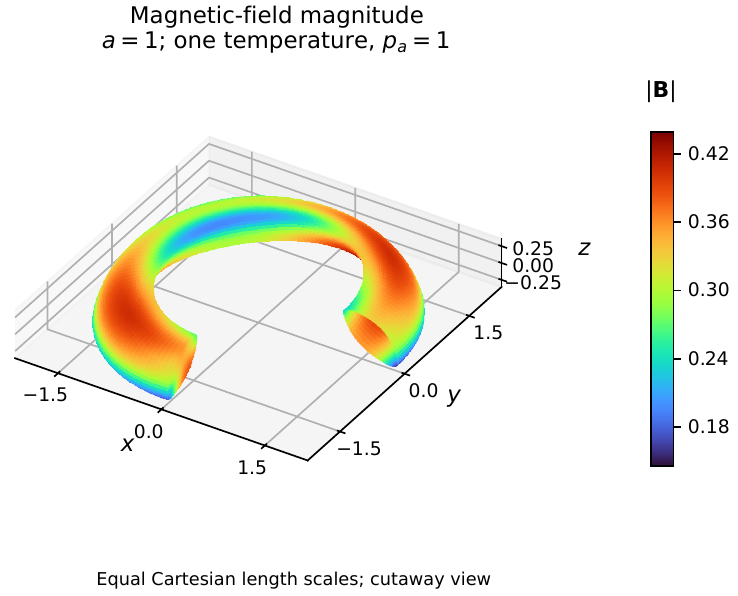}
\caption{Magnetic-field magnitude.}
\end{subfigure}\hfill
\begin{subfigure}{.49\linewidth}
\includegraphics[width=\linewidth]{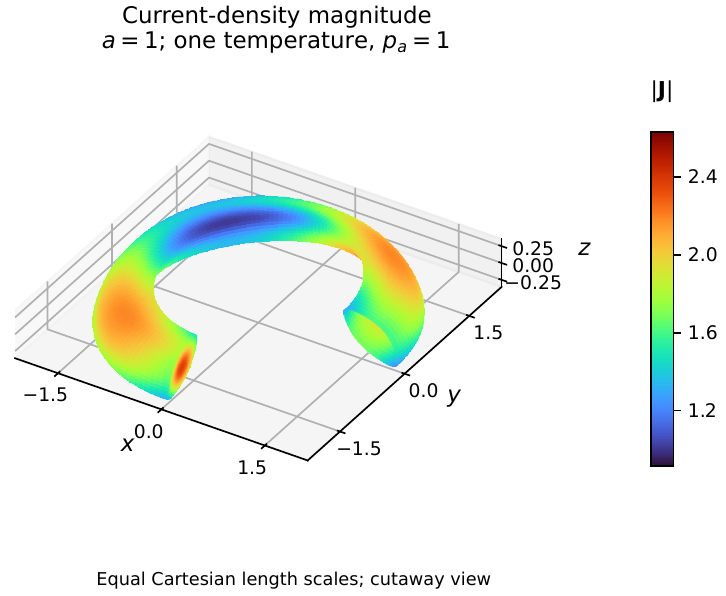}
\caption{Current-density magnitude.}
\end{subfigure}
\caption{Exact Cartesian magnetic and current fields on a cutaway view of the non-axisymmetric solid torus. A linear Turbo rainbow color scale encodes field values without illumination shading or height-based coloring. The exposed cuts reveal radial structure while the remaining exterior is the magnetic surface $m=m_b$.}
\label{fig:spatialBJ}
\end{figure}

\begin{figure}[tbp]
\centering
\begin{subfigure}{.49\linewidth}
\includegraphics[width=\linewidth]{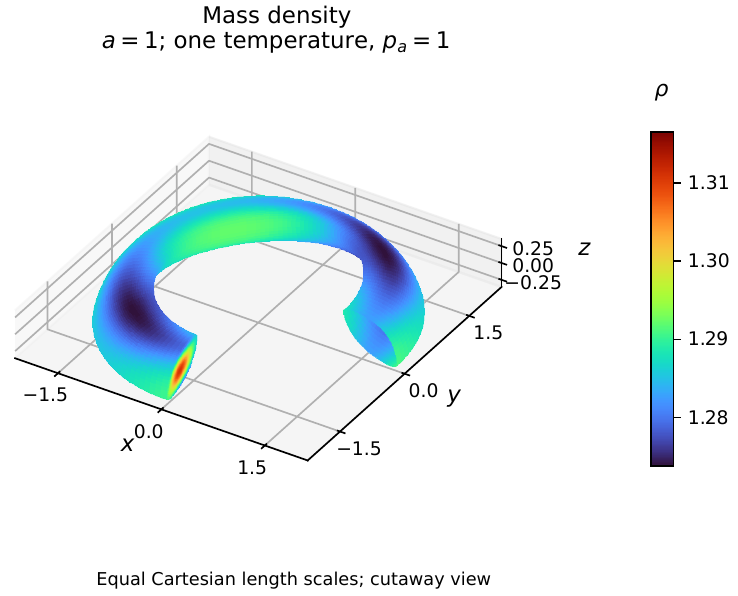}
\caption{Mass density.}
\end{subfigure}\hfill
\begin{subfigure}{.49\linewidth}
\includegraphics[width=\linewidth]{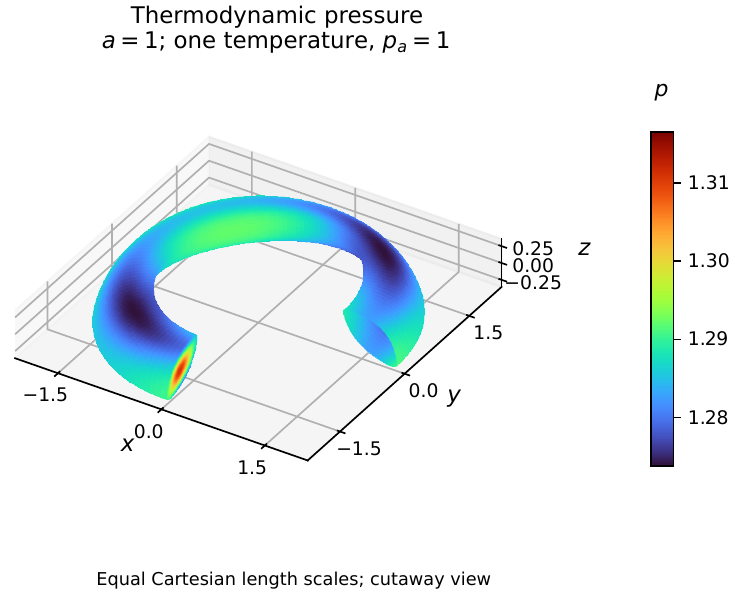}
\caption{Thermodynamic pressure.}
\end{subfigure}
\caption{Material-state distributions for the same exact field and time as Figure~\ref{fig:spatialBJ}. The colorbars intentionally resolve the small but nonzero variation. Equality of the numerical density and pressure values follows from $RT=1$ in this particular nondimensional state; it is not a general identity.}
\label{fig:spatialrhop}
\end{figure}

\begin{figure}[tbp]
\centering
\begin{subfigure}{.49\linewidth}
\includegraphics[width=\linewidth]{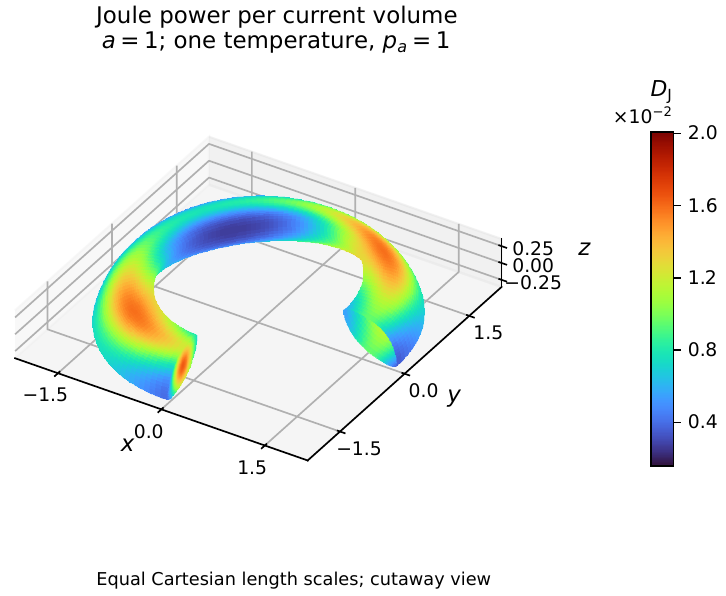}
\caption{Joule conversion.}
\end{subfigure}\hfill
\begin{subfigure}{.49\linewidth}
\includegraphics[width=\linewidth]{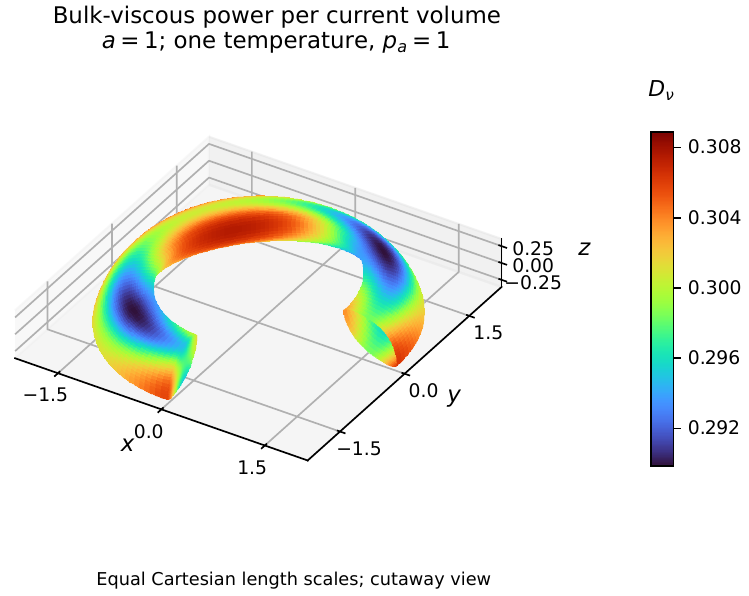}
\caption{Bulk-viscous conversion.}
\end{subfigure}
\caption{Spatially heterogeneous but strictly positive dissipative powers. Their opposite variations along a fixed magnetic surface follow from their explicit $g$-dependence. The distinct colorbar scales preserve the fact that bulk-viscous dissipation dominates this baseline.}
\label{fig:spatialpowers}
\end{figure}

Figures~\ref{fig:spatialBJ}--\ref{fig:spatialpowers} evaluate the Cartesian fields directly on the solid torus, rather than interpreting a time-profile curve as a spatial distribution. A quarter-sector is omitted from the rendering to expose two radial cuts through the interior. The cuts are visualization surfaces only; they neither change $\Omega_0$ nor introduce new boundary conditions. All three coordinate axes use the same physical length scale, and the colorbars report the actual nondimensional field values. The displayed baseline is the one-temperature solution at $a=1$; all maps use the same geometry and viewpoint.

The magnetic and current magnitudes in Figure~\ref{fig:spatialBJ} exhibit substantial variation both along and across the magnetic surfaces. On the sampled full domain, $|\BB|$ ranges approximately from $0.1460$ to $0.4392$ and $|\JJ|$ from $0.9163$ to $2.6299$. These are sampled ranges, not certified extrema. The current is not a constant scalar multiple of the field: its horizontal components contain the spatially varying matrix in \eqref{eq:current}. Accordingly, magnetic intensity alone does not determine the Joule-power distribution. The latter also depends on the inverse-designed resistivity tensor through $D_{\rm J}=\JJ\cdot\etab\JJ$.

Figure~\ref{fig:spatialrhop} resolves the weaker material-state heterogeneity. For this baseline, $RT=1$ at $a=1$, so the nondimensional pressure and density have the same numerical values, approximately $1.2738$--$1.3165$ over the sampled volume. Their similar appearance is an equation-of-state consequence, not a duplicated field. Neither field is generally constant on a magnetic surface, because its spatial factor contains $g$ as well as $p_0$. On a fixed magnetic surface, $p_0$ is constant while
\begin{equation}
p=\frac{b_*^2a^{-9/2}}9(12p_0-g),\qquad
\Pi=\frac{b_*^2a^{-9/2}}9(3p_0-g),\qquad
p-\Pi=b_*^2a^{-9/2}p_0.
\label{eq:fieldmap_pressure}
\end{equation}
Thus pressure and isotropic viscous stress have identical tangential variations, which cancel in the force-balancing effective pressure. The supplementary effective-pressure map is constant on the outer magnetic surface but varies across the exposed radial cuts. This distinguishes the present thermoviscous realization from interpreting $p_0$ itself as the evolving thermal pressure.

The two positive power densities in Figure~\ref{fig:spatialpowers} have different spatial patterns and markedly different absolute scales. At fixed $m$, $D_{\rm J}\propto g$ whereas $D_\nu\propto3p_0-g$; their tangential variations are consequently opposed. The sampled values are approximately $0.00159$--$0.02008$ for $D_{\rm J}$ and $0.28986$--$0.30888$ for $D_\nu$. The same Turbo palette is used with separate, explicitly labeled value ranges; similar colors in the two panels therefore do not imply similar heating magnitudes. Their integrated partition is quantified below by \eqref{eq:joulefraction}.

The geometrically corresponding maps at later times follow directly from the exact solution:
\begin{equation}
\begin{aligned}
|\BB|(a\XX,t)&=a^{-9/4}|\BB|(\XX,0),&
|\JJ|(a\XX,t)&=a^{-13/4}|\JJ|(\XX,0),\\
\rho(a\XX,t)&=a^{-3}\rho_0(\XX),&
p(a\XX,t)&=a^{-9/2}p(\XX,0),\\
D_{\rm J}(a\XX,t)&=a^{-11/2}D_{\rm J}(\XX,0),&
D_\nu(a\XX,t)&=a^{-11/2}D_\nu(\XX,0).
\end{aligned}
\label{eq:fieldmap_scalings}
\end{equation}
These relations preserve the material patterns but change their physical size and magnitude. In contrast, $T=T_*a^{-3/2}$ is spatially uniform at each time. The supplementary maps therefore use uniform temperature coloring and a common color and coordinate scale at $a=1$ and $a=2$. In particular, $\grad T=0$ and $\qv=0$ on this solution. Although the dissipative powers are nonuniform, their sum is proportional to the same material factor $12p_0-g$ as $\rho_0$, so the homogeneous temperature rate satisfies the local heat equation. Further three-dimensional maps show velocity, resistivity eigenvalue, bulk viscosity, reference-volume Joule history, entropy production and the two-temperature exchange fields.

\subsection{Magnetic geometry, phase dependence and the literature limit}
\begin{figure}[htbp]
\centering
\begin{subfigure}{.48\linewidth}\includegraphics[width=\linewidth]{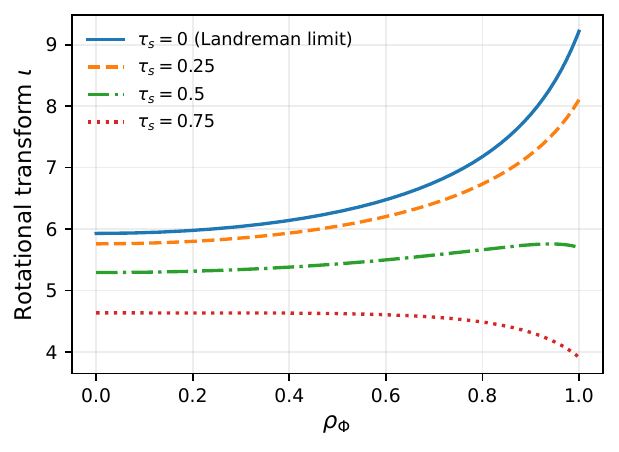}
\caption{Rotational transform.}\end{subfigure}\hfill
\begin{subfigure}{.48\linewidth}\includegraphics[width=\linewidth]{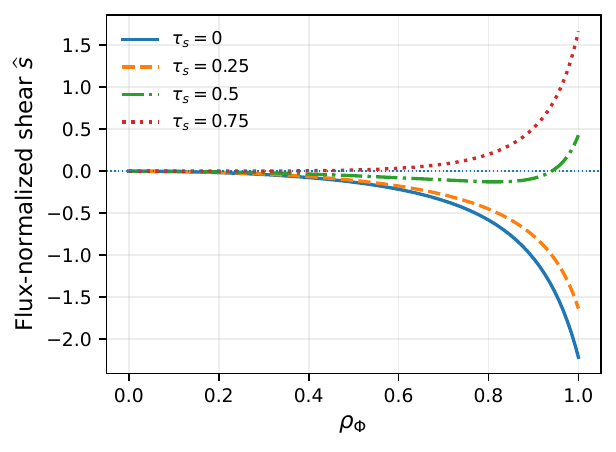}
\caption{Flux-normalized shear.}\end{subfigure}
\caption{Phase-dependent magnetic profiles from the exact period integrals, plotted against the flux radius of each configuration. The $\tau_s=0$ curve is the sheared Landreman reference field; nonzero phase gives the Issan extension. All configurations use $\epsilon=0.5$, $S=3$, $\lambda=4$ and $m_b=0.9$. The coordinate normalizes each configuration by its own boundary flux.}
\label{fig:transform}
\end{figure}

Figures~\ref{fig:geometry} and \ref{fig:expansion} show the actual reference torus and its homologous expansion. The field lines lie on embedded magnetic surfaces; their winding is not inferred from a visual projection. The phase study in Figure~\ref{fig:transform} evaluates the exact period formulas \eqref{eq:periods}. Only the one-dimensional integrals and the surface-to-flux coordinate are numerical. The derivative of the period ratio is evaluated by differentiating the integrands analytically, rather than by differentiating a fitted curve.

Increasing $\tau_s$ over the displayed range lowers the axis transform from approximately $5.92852$ at $\tau_s=0$ to $4.63571$ at $\tau_s=0.75$. It changes the radial profile more strongly than a uniform rescaling: the wall values decrease from $9.21831$ to $3.91271$, and an interior transform maximum appears in the displayed examples. For $\tau_s=0.5$, a zero of $\iota_m$ is located at $m\simeq0.823664$; for $\tau_s=0.75$, it moves to $m\simeq0.0384914$. The normalized shear changes sign at the same interior surfaces, with the opposite sign convention to $\iota_m$. These locations describe the evaluated parameter examples, not a proof of global uniqueness of extrema for all parameters.

The mechanism is visible in the separated periods: $\tau_s$ changes the hyperbolic factor in $\mathcal R(m)$, while $\mathcal P(m)$ is unchanged. Radial dependence prevents this change from reducing to a constant multiplier of $\iota$. By contrast, $\lambda$ changes the vertical length scale, magnetic flux magnitude and resistive rate but cancels from the transform ratio. This distinguishes a genuine change of winding from a change of geometry scale or traversal time, in the usual magnetic-surface interpretation \citep{Helander2014}.

The comparison to existing analytic equilibria is exact at the magnetic level. Setting $\tau_s=0$ in \eqref{eq:complexseed}--\eqref{eq:seedfield} recovers the sheared family of \citet{Landreman}; nonzero $\tau_s$ gives the extension of \citet{Issan}. On the evolving domain,
\begin{equation}
\frac{\BB(a\XX,t)}{b_*a^{-9/4}}=\BB_0(\XX),\qquad
\frac{p-\Pi}{b_*^2a^{-9/2}}=p_0(\XX).
\label{eq:literaturerecovery}
\end{equation}
Thus the normalized field and effective pressure recover the cited equilibrium at every time. The physical thermal pressure is generally different, as required by \eqref{eq:thermalpressure}. No regular zero-resistivity limit of the \emph{complete} thermoviscous model is claimed: in this family $\beta_0\propto\eta_*^{-1}$, so setting $\eta_*=0$ would be a singular constitutive operation. The magnetic reference comparison must not be presented as convergence to a classical ideal-plasma model.

\subsection{Cooling, resistive flux loss and the expansion clock}
At a fixed material scale $a$, both thermal branches have the same magnetic evolution. Material flux is $2^{-1/4}\simeq0.84090$ of its initial value at $a=2$ and $4^{-1/4}\simeq0.70711$ at $a=4$, while the winding and normalized shear remain unchanged. Resistive flux loss therefore does not require a change in the winding of this specially constructed field. It is the amplitude relative to the material area, rather than the appearance of the magnetic surfaces, that detects nonideal induction.

\begin{figure}[htbp]
\centering
\begin{subfigure}{.48\linewidth}\includegraphics[width=\linewidth]{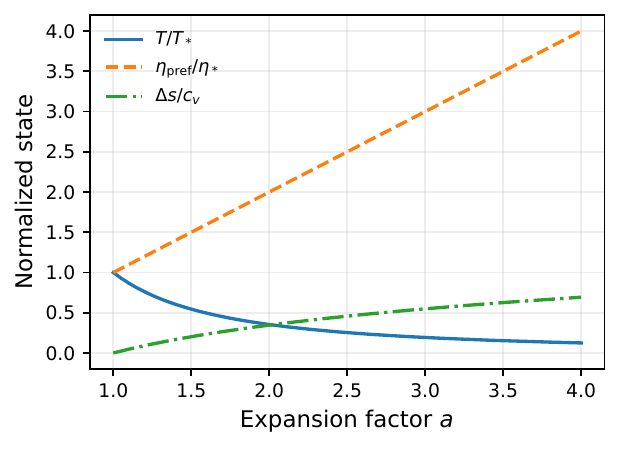}
\caption{Thermal state and resistivity.}\end{subfigure}\hfill
\begin{subfigure}{.48\linewidth}\includegraphics[width=\linewidth]{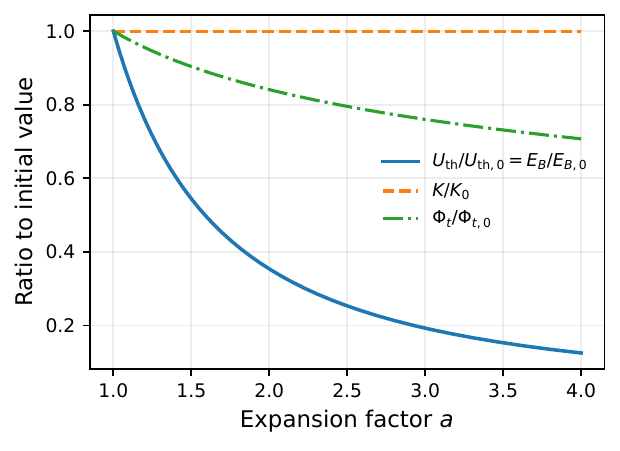}
\caption{Energy and flux ratios.}\end{subfigure}
\caption{Mechanisms along the one-temperature trajectory. Increasing resistivity coexists with cooling, positive entropy production and decreasing integrated magnetic and internal energies. The magnetic and thermal energy curves coincide after normalization because they share the same time exponent, not because the two energies are equal.}
\label{fig:thermalenergy}
\end{figure}

Figure~\ref{fig:thermalenergy} separates cooling from entropy production. Temperature falls as $a^{-3/2}$, whereas the resistivity prefactor grows as $a$. Nevertheless, $D_{\rm J}$ and $D_\nu$ both decrease as $a^{-11/2}$: current dilution and declining strain rate dominate the increasing resistivity. The reference internal-energy budget \eqref{eq:materialbudget} shows that accumulated pressure work is four times the accumulated irreversible heating. The net cooling is therefore consistent with positive entropy production. Magnetic and thermal energies integrated over the expanding volume both scale as $a^{-3/2}$, whereas material kinetic energy remains constant. This coupled interpretation is essential when comparing with dissipative exact solutions or thermodynamically consistent schemes \citep{Lesaffre2007,Montani,Gawlik2025}; magnetic diffusion alone would not determine the cooling law.

\begin{figure}[htbp]
\centering
\begin{subfigure}{.48\linewidth}\includegraphics[width=\linewidth]{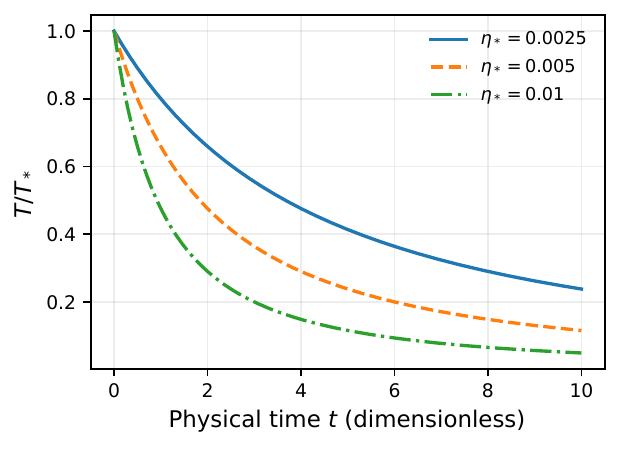}
\caption{Temperature.}\end{subfigure}\hfill
\begin{subfigure}{.48\linewidth}\includegraphics[width=\linewidth]{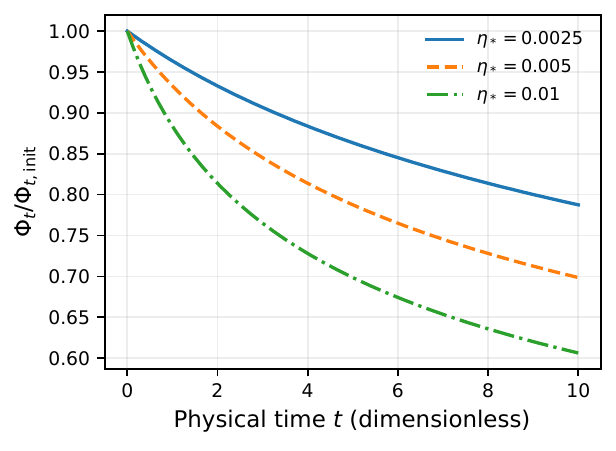}
\caption{Material magnetic flux.}\end{subfigure}
\caption{Effect of the reference resistive scale on physical-time evolution. Each curve is an elementary exact solution with its corresponding passive material texture. The time coordinate is dimensionless; identical values of $a$, rather than identical times, give identical normalized states.}
\label{fig:ratecomparison}
\end{figure}

Figure~\ref{fig:ratecomparison} varies $\eta_*$ at fixed physical time. Since $a=1+4\lambda^2\eta_*t$, larger $\eta_*$ accelerates expansion, cooling and material-flux loss. At $t=5$, the temperature ratios for $\eta_*=0.0025,0.005,0.01$ are approximately $0.41409,0.23853,0.11618$, and the flux ratios are $0.86334,0.78751,0.69853$. When plotted against $a$, these curves collapse onto the same exact state laws. This family varies the dissipative clock, not the material shear profile. Because $U$ and $\beta_0$ are also changed consistently, the comparison is not an experiment in which only electrical resistivity changes while all other material properties and initial velocities remain fixed.

\subsection{Pressure offset and the partition of irreversible heat}
\begin{figure}[htbp]
\centering
\includegraphics[width=.49\linewidth]{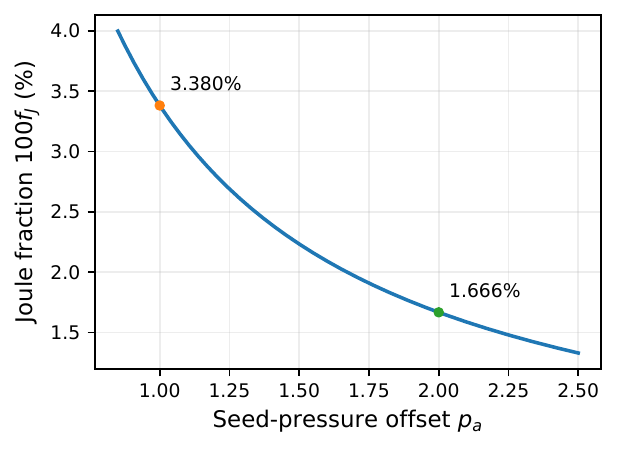}
\caption{Pressure-offset dependence of the heat partition at fixed magnetic geometry. The points mark the one- and two-temperature baseline offsets. The decline follows analytically from \eqref{eq:partitionderivative}; changing the offset changes the inverse-designed material, not its magnetic winding.}
\label{fig:partition}
\end{figure}

The volume-integrated Joule fraction has a closed expression because both dissipative powers share a time factor:
\begin{equation}
f_{\rm J}=\frac{\int D_{\rm J}\dd V}{\int(D_{\rm J}+D_\nu)\dd V}
=\frac{3I_g}{12I_p-I_g}.
\label{eq:joulefraction}
\end{equation}
For fixed geometry, let $V_0=|\Omega_0|$ and $I_\psi=\int_{\Omega_0}\psi\dd V$. Then $I_p=V_0p_a-I_\psi/\lambda^2$, while $I_g$ is independent of $p_a$. Therefore
\begin{equation}
\frac{\dd f_{\rm J}}{\dd p_a}
=-\frac{36I_gV_0}{(12I_p-I_g)^2}<0.
\label{eq:partitionderivative}
\end{equation}
Increasing the pressure offset increases the required density and bulk-viscosity texture but leaves the magnetic field, current and shear unchanged. Joule power is unchanged, while viscous heating increases. Figure~\ref{fig:partition} shows the resulting decrease of $f_{\rm J}$ from about $3.997\%$ at $p_a=0.85$ to $1.329\%$ at $p_a=2.5$; the baseline values are $3.380\%$ and $1.666\%$. These are fractions of irreversible conversion, not fractions of the net temperature change. Interspecies exchange is not a third total-energy source.
The dominance of bulk-viscous heating in these examples is a property of the selected realization. It prevents interpreting them as Joule-heating-dominated confinement scenarios. More conventional parallel/perpendicular resistive equilibria and collision-derived laws address different constitutive questions \citep{Poulipoulis,Lingam,Braginskii1965}; passivity of the present tensor alone does not identify it with those models.

\subsection{Two-temperature separation, entropy and finite validity}
\begin{figure}[htbp]
\centering
\begin{subfigure}{.48\linewidth}\includegraphics[width=\linewidth]{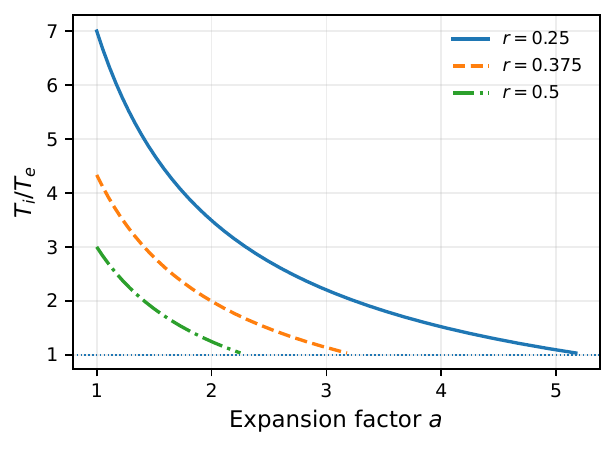}
\caption{Ion/electron temperature ratio.}\end{subfigure}\hfill
\begin{subfigure}{.48\linewidth}\includegraphics[width=\linewidth]{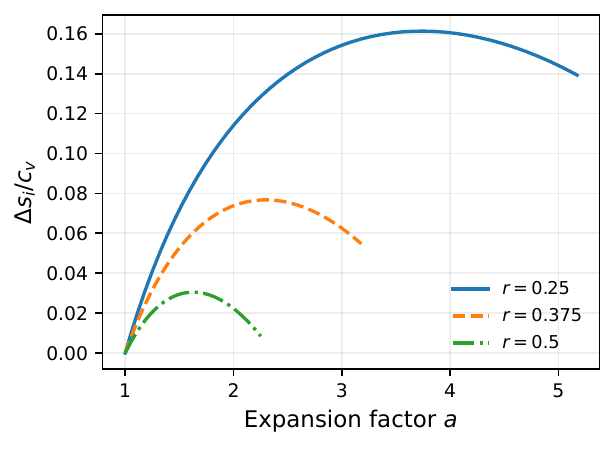}
\caption{Ion specific entropy.}\end{subfigure}
\caption{Dependence on the initial electron fraction at $p_a=2$. Each curve ends at $0.98a_{\rm eq}$, strictly before its own constitutive singularity. The decline in ion entropy represents outgoing heat exchange; the total production remains positive.}
\label{fig:twothermal}
\end{figure}

\begin{figure}[htbp]
\centering
\includegraphics[width=.49\linewidth]{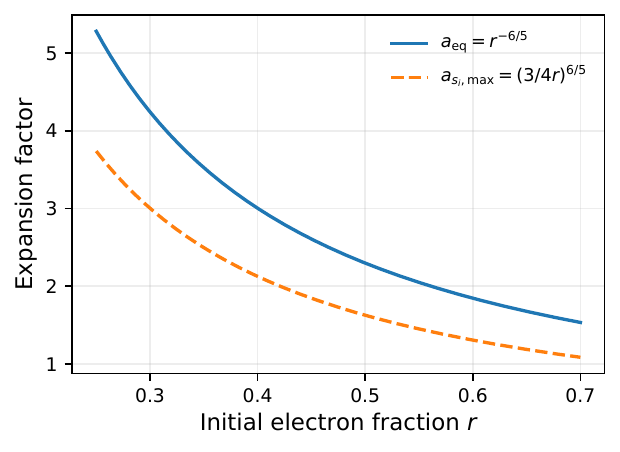}
\caption{Exact equality endpoint and ion-entropy maximum as functions of $r$. The upper curve is an excluded endpoint, not a stable final state or a numerically resolved continuation through the pole.}
\label{fig:interval}
\end{figure}

The initial electron fraction changes the temperature separation without changing the mean-temperature or magnetic trajectory. In particular,
\begin{equation}
\frac{T_i}{T_e}=\frac{2}{r}a^{-5/6}-1,\qquad
a_{\rm eq}=r^{-6/5},\qquad
\frac{a_{s_i,\max}}{a_{\rm eq}}=(3/4)^{6/5}.
\label{eq:rcomparison}
\end{equation}
Increasing $r$ moves equality earlier. For $r=0.25,0.375,0.5$, the endpoint factors are respectively $5.27803,3.24461,2.29740$. The ion-entropy maxima occur at $3.73719,2.29740,1.62671$. Figure~\ref{fig:twothermal} shows how the temperature ratio approaches unity while ion entropy first rises and then decreases; Figure~\ref{fig:interval} displays the shrinking admissible interval. The corresponding total entropy remains increasing by \eqref{eq:twoentropy}. Cooling alone does not determine either species' entropy change: density dilution and the direction of interspecies transfer also enter.
At equality the required exchange power is still strictly positive, as established by \eqref{eq:Qendpoint}--\eqref{eq:Gpole}. A bounded temperature-difference law could not maintain this path. The result therefore characterizes the finite scope of the inverse-designed two-temperature realization, rather than predicting a singular Coulomb equilibration process. The $T_e^{-3/2}$ scaling agrees with the named temperature prefactor, but the tensor and exchange coefficient remain different from classical collisional transport \citep{Spitzer,Braginskii1965}.

\subsection{Boundary energy transfer and physical interpretation}
\begin{figure}[htbp]
\centering
\begin{subfigure}{.48\linewidth}\includegraphics[width=\linewidth]{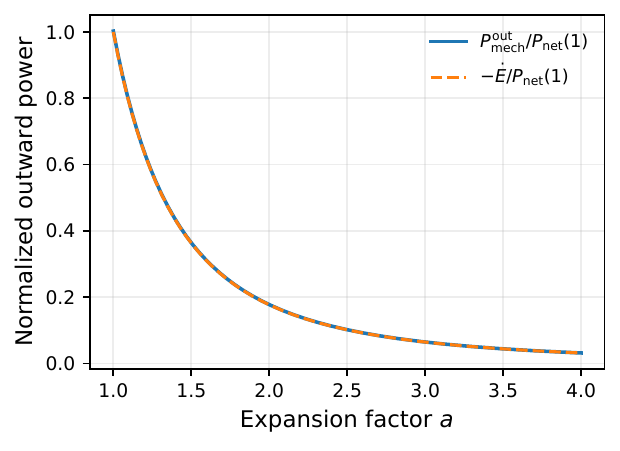}
\caption{Mechanical and net outward power.}\end{subfigure}\hfill
\begin{subfigure}{.48\linewidth}\includegraphics[width=\linewidth]{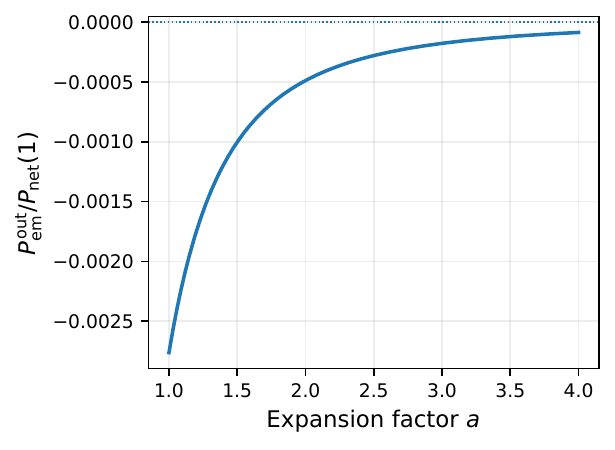}
\caption{Relative electromagnetic power.}\end{subfigure}
\caption{Moving-boundary energy balance, normalized by $P_{\rm net}(1)=-\dot E(0)$. Mechanical work removes energy, whereas the electromagnetic port supplies a smaller amount. The curves share an $a^{-5/2}$ factor. Energy closure is a complete port balance, not constant energy in an isolated torus.}
\label{fig:boundary}
\end{figure}

For the one-temperature reference geometry, independent volume quadrature gives
\begin{equation}
\begin{aligned}
|\Omega_0|&\simeq3.1501412006,&I_p&\simeq3.0977071246,\\
I_g&\simeq0.4141467264,&I_B&\simeq0.3100467783.
\end{aligned}
\label{eq:quadraturenumbers}
\end{equation}
Since $I_B-I_g<0$, the relative Poynting power is inward. At $a=1$, the mechanical outward power is approximately $3.0234063$, the electromagnetic contribution is $-0.0083280$, and the net outward power is $3.0150783$, equal to $-\dot E$. The electromagnetic input is only $0.2762\%$ of the net outward power in magnitude; plotting both contributions on a single scale obscures its sign. Figure~\ref{fig:boundary} therefore uses separate panels with a common normalization.
The pressure and magnetic stresses do work on the expanding boundary even though material acceleration is zero. A nonzero relative electric trace supplies electromagnetic power while material flux decreases. Neither feature contradicts positive bulk dissipation: boundary exchange and internal conversion are different terms. This distinction is the same one emphasized by bulk--boundary and inter-channel power formulations \citep{PortHamiltonian,RenMultiChannel}, and it is important for testing moving-mesh MHD implementations. The present exact boundary data do not replace a solved vacuum/coil/wall system.

\subsection{Independent verification and reproducibility}
Symbolic calculations verify the time exponents, power and history identities, and positive-minor expressions of the resistivity tensor. Cartesian automatic differentiation evaluates the current independently of its explicit formula. A second implementation uses high-precision arithmetic and obtains the electric field from \eqref{eq:potentialpair} before evaluating the Ohm-law residual. Table~\ref{tab:residuals} summarizes the differential checks.
\begin{table}[htbp]
\centering\small
\caption{Differential residuals. The reported maxima are taken over the sampled points and do not constitute uniform error bounds.}
\label{tab:residuals}
\begin{tabularx}{\linewidth}{@{}p{.23\linewidth}Xr@{}}\toprule
Check & Method and coverage & Largest residual\\\midrule
Reference field & Cartesian derivatives at 48 points & $3.56\times10^{-15}$\\
One-temperature system & 192 space--time cases & $8.89\times10^{-16}$\\
Two-temperature system & 72 space--time cases & $2.67\times10^{-15}$\\
Independent implementation & 55-decimal arithmetic; 12 cases & $8.2\times10^{-56}$\\
Scalar obstruction & Symbolic third-order Cartesian expansion & Exact identity\\\bottomrule
\end{tabularx}
\end{table}
\FloatBarrier
The momentum residual tests force cancellation through $p-\Pi$, the thermal residual tests cooling against pressure work and heating, and the history residual tests material transport of accumulated power. The multiprecision residuals measure agreement between independent implementations of the analytical formulas. They do not quantify physical-model accuracy or establish stability.

For the parameter studies, the period integrals and their analytic derivatives were evaluated at multiple angular resolutions and compared with adaptive quadrature. The largest refinement difference at the control points was $1.8\times10^{-15}$; a separate finite-difference derivative check was below $1.7\times10^{-8}$. Independent volume quadratures for the pressure-offset analysis differed by less than $1.1\times10^{-11}$. These calculations evaluate the exact solution formulas by quadrature, without numerical time integration of the MHD equations. The moving-boundary surface--volume discrepancies decrease from $4.71\times10^{-10}$ to $9.2\times10^{-15}$ and $6.9\times10^{-16}$ with quadrature refinement. The checked evaluator additionally rejects nonfinite data, invalid toroidal branches, negative viscosity and states at or beyond the two-temperature endpoint.

\subsection{Formal verification}\label{sec:formal}
The accompanying Lean~4.19.0 formalization \citep{LeanValidation} comprises 407 theorems and five interface checks in 37 modules. Compilation, declaration-type extraction and axiom auditing cover 626 named declarations. Their logical dependencies are confined to \texttt{Classical.choice}, \texttt{Quot.sound} and \texttt{propext}. The source code, build records and declaration-level audits are maintained at \repositoryurl, with the proof archive identified by \repositoryarchive. A correspondence table relates the Lean declarations and their hypotheses to stable LaTeX labels.

Layer~1 contains finite-dimensional passivity and balance identities. Layer~2 formalizes a single-degree-of-freedom thermo-viscoelastic reduction of Section~\ref{sec:review}; Layer~3 treats single-temperature, magnetic-quasistatic Eulerian MHD with independent vector-potential variations and fixed-box integration. Cross-layer results retain their normalization, reference-state and interface-power hypotheses. The Revision3 modules cover a nonmagnetic specialization and scalar identities on the prescribed expansion branch, including the cumulative pressure-work/irreversible-heat ratio and the derivative, maximum and monotonicity intervals of the current-volume Joule history.

The formalization does not include the full Maxwell equations, the two-temperature PDE system, toroidal embedding, a complete three-dimensional realization of the homologous branch, or the moving-boundary energy theorem. The analytical results for the toroidal and two-temperature solutions therefore extend beyond these formalized specializations. The scope and declaration counts of each formal module are listed in the supplementary material.


\section{Conclusions}\label{sec:discussion}
A passive anisotropic resistivity tensor extends a sheared non-axisymmetric toroidal equilibrium to exact thermoviscous evolution within the PDEL formulation. The field retains its spatial structure while material magnetic flux decays. Momentum and thermal balance determine the density, bulk viscosity and temperature scaling. The one-temperature branch is defined at every finite forward time; the two-temperature branch admits a $T_e^{-3/2}$ resistivity prefactor on a finite interval.

The solutions preserve magnetic winding during resistive flux loss and exhibit cooling with positive irreversible entropy production. Their transported power histories and moving-boundary balance account for expansion work, internal conversion and electromagnetic exchange through the same constitutive responses. They therefore provide benchmarks for coupled magnetic, mechanical and thermal calculations. Within PDEL, the weak field equation, realized power and thermal receiver are linked by a single constitutive action.

The coefficients define a passive heterogeneous continuum rather than a classical collisional-plasma closure. The scalar-resistivity obstruction and the two-temperature exchange pole restrict the admissible trajectories. Magnetic-shear redistribution, perturbation stability and coupling to an exterior vacuum--coil--wall system remain open extensions.


\section*{Data and code availability}
The analytical field evaluator, symbolic derivations, parameter-study data and residual calculations accompany the manuscript. The Lean formalization and verification records are maintained at \repositoryurl; \repositoryarchive\ identifies the archived proof sources and audits. Access to the repository and its release files is currently restricted. The mathematical scope of the formalization is specified in Section~\ref{sec:formal}. The numerical code evaluates the exact solutions and their residuals.
\appendix
\section{Toroidal coordinates and the shear calculation}\label{app:geometry}
The Cartesian field is the sheared family of \citet{Landreman} with the complex-phase extension of \citet{Issan}. The parameterization below makes its regular magnetic surfaces and period integrals explicit for the present material-domain calculation; it is not a new static equilibrium family. For the seed \eqref{eq:seedfield}, set $m=2\psi$, $k=\sqrt m$ and define
\begin{equation}
\begin{aligned}
n(\zeta)&=\tau_s+\frac\epsilon2\sin2\zeta,&
\sigma(\theta)&=S+\arcsin(k\sin\theta),\\
h_g&=\sqrt{4\sigma^2+\epsilon^2},&
G&=\frac{h_g+\epsilon\cos2\zeta}{2},\\
f&=\sqrt{1-m\sin^2\theta},&
L&=\sqrt{\cosh^2n-m},\\
a_c&=\sqrt{(h_g-\epsilon)/2},&
b_c&=\sqrt{(h_g+\epsilon)/2}.
\end{aligned}
\label{eq:mapfunctions}
\end{equation}
The surface map is
\begin{equation}
\boldsymbol r(m,\theta,\zeta)=
\left(a_c\cos\zeta,\ b_c\sin\zeta,
\frac1\lambda\left[
\arctan\frac{k\cos\theta}{L}
-\arctan\frac{k\sin\theta\tanh n}{f}\right]\right).
\label{eq:surfacemap}
\end{equation}
Both angles are $2\pi$-periodic and both arctangents use the principal branch. Under \eqref{eq:regularity}, $\sigma\ge S-\arcsin\sqrt{m_b}>0$, so $a_c,b_c,G,f,L$ remain positive. The horizontal coordinates satisfy
\begin{equation}
a_cb_c=\sigma,\qquad b_c^2-a_c^2=\epsilon,\qquad
K=b_c\cos\zeta-i a_c\sin\zeta,\qquad
wK=\sigma+i\frac\epsilon2\sin2\zeta.
\label{eq:ellipticidentities}
\end{equation}
They are single-valued elliptic coordinates outside the excluded focal segment.

For a fixed horizontal point, introduce the disk variables
\[
u_d=\sin(\sigma-S)=k\sin\theta,
\qquad v_d=-\Im(e^{-i\lambda Z}\sin\Xi)=k\cos\theta.
\]
They satisfy $u_d^2+v_d^2=m$ and $e^{-i\lambda Z}\sin\Xi=L-ik\cos\theta$. On the selected branch, $\partial_Zv_d=\lambda L>0$, so the vertical inverse is unique; \eqref{eq:surfacemap} gives that inverse. It lies within $|\lambda Z|\le\arcsin\sqrt{m_b}<\pi/2$. All functions are smooth in $(u_d,v_d)$, including at $m=0$, where only the polar labels degenerate. The magnetic axis is the planar ellipse obtained by setting $m=0$.

Direct differentiation yields the volume element and contravariant field components
\begin{equation}
\dd V=\frac{G}{\lambda h_gfL}\dd\psi\dd\theta\dd\zeta,
\qquad B_0^\theta=\frac{h_gf}{2G}>0,
\qquad B_0^\zeta=\frac{L}{2G}>0.
\label{eq:jacobian}
\end{equation}
Together with the inverse argument, these relations establish the embedded solid torus and the nonvanishing field.

The separated field-line construction follows the analytic seed geometry \citep{Landreman,Issan}. The two-angle period ratio is consistent with the rotational-transform convention in toroidal confinement theory \citep{Kruskal1958,Helander2014}. The field-line equation separates:
\begin{equation}
\frac{\dd\theta}{h_g\sqrt{1-m\sin^2\theta}}
=\frac{\dd\zeta}{\sqrt{\cosh^2[\tau_s+\epsilon\sin2\zeta/2]-m}}.
\label{eq:fieldlineseparation}
\end{equation}
Its complete periods are
\begin{equation}
\begin{aligned}
\mathcal P(m)&=\int_0^{2\pi}
\frac{\dd\theta}
{\sqrt{4[S+\arcsin(\sqrt m\sin\theta)]^2+\epsilon^2}
\sqrt{1-m\sin^2\theta}},\\
\mathcal R(m)&=\int_0^{2\pi}
\frac{\dd\zeta}{\sqrt{\cosh^2[\tau_s+\epsilon\sin2\zeta/2]-m}}.
\end{aligned}
\label{eq:periods}
\end{equation}
The transform in \eqref{eq:transform} follows by counting the two angular periods. For a fixed $\zeta$ cut, multiplication of the Jacobian by $B_0^\zeta$ gives $\dd\Phi_{t,0}/\dd\psi=\mathcal P(2\psi)/(2\lambda)$, which proves the flux and shear formulas.

For the near-axis shear, write $N_j=\langle\sech^j n\rangle$ and $f_0(s)=(4s^2+\epsilon^2)^{-1/2}$. The period averages expand as
\[
\frac{\mathcal R(m)}{2\pi}=N_1+\frac m2N_3+O(m^2),\qquad
\frac{\mathcal P(m)}{2\pi}=f_0(S)+\frac m4[f_0(S)+f_0''(S)]+O(m^2).
\]
Hence
\begin{equation}
\frac{\iota_\psi(0)}{\iota(0)}
=\frac{N_3}{N_1}-\frac12-
\frac{2(8S^2-\epsilon^2)}{(4S^2+\epsilon^2)^2}.
\label{eq:axialshear}
\end{equation}
To bound the first term, the weight $\sech\sqrt z$ decreases with $z=n^2$. Its covariance with $z$ is therefore nonpositive, and $\tanh^2n\le n^2$ gives
\[
\frac{N_3}{N_1}=1-\mathbb E_{\sech n}(\tanh^2n)
\ge1-\tau_s^2-\frac{\epsilon^2}{8}.
\]
At $\epsilon=\tau_s=1/2$ and $S=3$, the first term is at least $23/32$ and the final geometric term is strictly less than $1/9$. Thus $\iota_\psi(0)/\iota(0)>23/32-1/2-1/9=31/288$, as used in the main text. The seed geometry and field-line structure are attributed to \citet{Landreman,Issan}; the present calculation verifies the parameter choice used for the dissipative realization.

\section{Axis calculation for the scalar-resistivity obstruction}\label{app:scalar}
Scalar and anisotropic resistive closures lead to different compatibility restrictions \citep{Poulipoulis,Montani}. The following calculation is specific to the attributed field \eqref{eq:seedfield} \citep{Landreman,Issan}; those references are the source of the magnetic field, not a prior proof of the obstruction derived here. Let $\boldsymbol K_0=\grad_X\!\times\JJ_0$. If \eqref{eq:scalareigen} held, its scalar product with $\JJ_0$ would imply
\begin{equation}
r\,\JJ_0\cdot\boldsymbol K_0=\Lambda\,\JJ_0\cdot\BB_0.
\label{eq:scalardot}
\end{equation}
At the outer magnetic-axis point $(x_a,0,0)$ define
\begin{equation}
x_a^2=\frac{\sqrt{4S^2+\epsilon^2}-\epsilon}{2},\qquad
k_a=\sqrt{x_a^2+\epsilon},\qquad
x_ak_a=S,\qquad D_a=\lambda^2+4x_a^2.
\label{eq:axisparameters}
\end{equation}
Both products on the right and left of \eqref{eq:scalardot}, excluding $r$ and $\Lambda$, are positive there. Thus $\Lambda>0$, and in a neighborhood the only candidate is
\begin{equation}
r_s:=\frac r\Lambda
=\frac{\JJ_0\cdot\BB_0}{\JJ_0\cdot\boldsymbol K_0}.
\label{eq:uniquescalar}
\end{equation}
Differentiating the full Cartesian expressions \eqref{eq:seedfield} and \eqref{eq:current}, and only then evaluating on the axis, gives
\begin{equation}
\begin{aligned}
r_s&=D_a^{-1},&
(r_s)_X&=-\frac{8x_a\lambda^2}{(D_a+2\epsilon)D_a^2},\\
J_{0y}&=\frac{D_a+2\epsilon}{2\lambda k_a}\cosh\tau_s,&
K_{0z}&=\frac{4x_a}{\lambda k_a}\cosh\tau_s,
\end{aligned}
\label{eq:axisderivatives}
\end{equation}
with $J_{0x}=B_{0z}=0$. Therefore
\begin{equation}
\left.\left[\grad_X\!\times(r_s\JJ_0)-\BB_0\right]\right|_{(x_a,0,0)}
=\left(-\frac{8x_a\sinh\tau_s}{D_a^2},\ 0,\
\frac{16x_a^3\cosh\tau_s}{\lambda k_aD_a^2}\right).
\label{eq:fullaxisresidual}
\end{equation}
The third component is strictly positive, proving Proposition~\ref{prop:scalar}.

The derivative in \eqref{eq:axisderivatives} cannot be obtained by differentiating the axis-only value $r_s=D_a^{-1}$ as if it were an off-axis formula. In particular, differentiation of $\JJ_0\cdot\boldsymbol K_0$ includes $J_{0z,X}K_{0z}$. This term is retained in the independent third-order Taylor calculation supplied with the implementation. For Table~\ref{tab:parameters}, the third residual component is approximately $0.01566689095$.

\section{Homologous exponent matching and passivity}\label{app:exponents}
Matching induction alone does not ensure thermal compatibility, as illustrated in other resistive exact-solution studies \citep{Lesaffre2007,Montani}. Here all exponents are determined from the particular PDEL-generated system in Section~\ref{sec:pdel}. Consider the restricted separation class
\begin{equation}
\begin{aligned}
a&=1+Ut,&T&=T_*a^{s_T},&
\rho&=a^{-3}\rho_0,\\
\BB&=b_*a^{\delta_B-2}\BB_0,&
\etab&=\eta_*(T/T_*)^\nu\MM_0,&
\zeta&=\beta_0\rho(T/T_*)^{q_\zeta},
\end{aligned}
\label{eq:generalansatz}
\end{equation}
with $U\ne0$ and nonzero resistive action. The symbol $\delta_B$ is a magnetic-amplitude exponent, distinct from the material viscosity texture $\beta_0$.

Using \eqref{eq:eigenidentity}, induction requires
\begin{equation}
\nu s_T=1,\qquad U\delta_B=-\kappa_0.
\label{eq:inductionmatching}
\end{equation}
Pressure and magnetic stress share an exponent if $s_T-3=2\delta_B-4$, whereas viscous stress shares it if $-4+q_\zeta s_T=2\delta_B-4$. Consequently, for $\nu\notin\{0,-1\}$,
\begin{equation}
s_T=\frac1\nu,\qquad
\delta_B=\frac{\nu+1}{2\nu},\qquad
U=-\frac{2\nu\kappa_0}{\nu+1},\qquad
q_\zeta=\nu+1.
\label{eq:matchingresult}
\end{equation}
Writing $\Pi=\Pi_0a^{s_T-3}$, momentum fixes
$\Pi_0=\rho_0RT_*-b_*^2p_0$. The thermal equation then gives
\begin{equation}
\begin{aligned}
\rho_0&=-\frac{b_*^2}{c_vT_*}
\left[\frac{\nu+1}{2}g+3\nu p_0\right],\\
\Pi_0&=-b_*^2\left[
\frac{(\gamma-1)(\nu+1)}2g+
\{1+3\nu(\gamma-1)\}p_0\right],\\
\beta_0&=\frac{\Pi_0}{3U\rho_0}.
\end{aligned}
\label{eq:generalmaterial}
\end{equation}
These formulas reproduce \eqref{eq:materialdata} at $\nu=-2/3$, $\gamma=5/3$.

For expansion, $U>0$ forces $-1<\nu<0$. With $p_0,g>0$, a sufficiently large positive pressure offset gives $\rho_0>0$, while $\Pi_0>0$ additionally requires $1+3\nu(\gamma-1)<0$. This yields \eqref{eq:passivewindow}; at its upper boundary the remaining $g$-term makes $\Pi_0<0$. The interval is nonempty only for $\gamma>4/3$.

At $\nu=-3/2$ and $\gamma=5/3$, \eqref{eq:matchingresult} instead gives $U=-6\kappa_0$, while \eqref{eq:generalmaterial} gives $\Pi_0=b_*^2(2p_0+g/6)>0$. Positive density would therefore require $\beta_0<0$. This calculation excludes that exponent from the present passive one-temperature homologous class, not from all possible thermoviscous evolutions. The two-temperature branch changes the pressure--temperature matching and so lies outside this one-temperature obstruction.

\section{Continuum thermo-viscoelastic first variation and energy balance}\label{app:review}
The reference-volume model of Section~\ref{sec:review} follows the thermo-mechanical PDEL state and power conventions \citep{RenEntropy,RenSpending,RenEulerian,RenMultiChannel}. This appendix displays the continuum calculation without replacing compatible displacement variations by scalar trajectory tests. Put $\boldsymbol\sigma=\boldsymbol\sigma_e+\boldsymbol\sigma_v$ and $\delta\boldsymbol\varepsilon=\sym\grad_X\delta\boldsymbol u$. The mixed state differential is
\begin{equation}
\widehat\delta(\psi+\theta s)
=\boldsymbol\sigma_e:\delta\boldsymbol\varepsilon
+(\psi_{,\theta}+s)\delta\theta
+\theta\dot s\,\delta t_s.
\label{eq:reviewmixedstate}
\end{equation}
Both histories in \eqref{eq:reviewhistories} have their past integrands locked. Combining \eqref{eq:reviewmixedstate} with \eqref{eq:reviewrules}, before integration by parts, gives
\begin{equation}
\begin{aligned}
\mathfrak D\mathscr A_{\rm tv}
={}&\int_{t_0}^{t_1}\!\int_{\Omega_0}
\Big[\rho_0\dot{\boldsymbol u}\cdot\delta\dot{\boldsymbol u}
-\boldsymbol\sigma:\sym\grad_X\delta\boldsymbol u
-(\psi_{,\theta}+s)\delta\theta\\
&\qquad-\big(\theta\dot s+\grad_X\!\cdot\boldsymbol Q-D_v\big)\delta t_s
+\boldsymbol b_0\cdot\delta\boldsymbol u\Big]\dd V_0\dd t\\
&+\int_{t_0}^{t_1}\!\int_{\Gamma_t}
\overline{\boldsymbol t}\cdot\delta\boldsymbol u\,\dd S_0\dd t.
\end{aligned}
\label{eq:reviewbeforeIBP}
\end{equation}
Time-endpoint vanishing of $\delta\boldsymbol u$ removes the kinetic endpoint term. Stress symmetry gives $\boldsymbol\sigma:\sym\grad_X\delta\boldsymbol u=\boldsymbol\sigma:\grad_X\delta\boldsymbol u$. The spatial divergence theorem yields
\begin{equation}
-\int_{\Omega_0}\boldsymbol\sigma:\grad_X\delta\boldsymbol u\,\dd V_0
=\int_{\Omega_0}(\grad_X\!\cdot\boldsymbol\sigma)\cdot\delta\boldsymbol u\,\dd V_0
-\int_{\partial\Omega_0}(\boldsymbol\sigma\boldsymbol N)\cdot\delta\boldsymbol u\,\dd S_0.
\label{eq:reviewstressIBP}
\end{equation}
These steps give \eqref{eq:reviewfirstvariation}, its bulk equations and the natural traction. No entropy conjugacy was imposed in evaluating the first variation.

After obtaining $s=-\psi_{,\theta}$, differentiation of $e_0=\psi+\theta s$ and use of the entropy balance give
\begin{equation}
\dot e_0
=\boldsymbol\sigma_e:\dot{\boldsymbol\varepsilon}+\theta\dot s
=\boldsymbol\sigma_e:\dot{\boldsymbol\varepsilon}
+\boldsymbol\sigma_v:\dot{\boldsymbol\varepsilon}
-\grad_X\!\cdot\boldsymbol Q.
\label{eq:reviewinternal}
\end{equation}
The kinetic energy rate follows independently from momentum:
\begin{equation}
\partial_t\big(\tfrac12\rho_0|\dot{\boldsymbol u}|^2\big)
=\grad_X\!\cdot(\boldsymbol\sigma^T\dot{\boldsymbol u})
-\boldsymbol\sigma:\dot{\boldsymbol\varepsilon}
+\boldsymbol b_0\cdot\dot{\boldsymbol u}.
\label{eq:reviewkinetic}
\end{equation}
Adding \eqref{eq:reviewinternal} and \eqref{eq:reviewkinetic} proves \eqref{eq:reviewenergy}, and integration over the fixed reference body gives
\begin{equation}
\begin{aligned}
\frac{\dd}{\dd t}\int_{\Omega_0}
\big(\tfrac12\rho_0|\dot{\boldsymbol u}|^2+e_0\big)\dd V_0
={}&\int_{\Omega_0}\boldsymbol b_0\cdot\dot{\boldsymbol u}\,\dd V_0\\
&+\int_{\partial\Omega_0}
(\boldsymbol\sigma\boldsymbol N)\cdot\dot{\boldsymbol u}\,\dd S_0
-\int_{\partial\Omega_0}\boldsymbol Q\cdot\boldsymbol N\,\dd S_0.
\end{aligned}
\label{eq:reviewboundarybudget}
\end{equation}
On $\Gamma_t$ the traction is prescribed; on a moving prescribed-displacement boundary the corresponding reaction work is retained. The histories do not add another storage term to this physical energy.

Finally, the entropy-flux form distinguishes thermal transport from production:
\begin{equation}
\dot s+\grad_X\!\cdot\left(\frac{\boldsymbol Q}{\theta}\right)
=\frac{D_v}{\theta}-\frac{\boldsymbol Q\cdot\grad_X\theta}{\theta^2}
=\frac{\dot{\boldsymbol\varepsilon}:\mathbb V:\dot{\boldsymbol\varepsilon}}{\theta}
+\frac{\grad_X\theta\cdot\boldsymbol K_0\grad_X\theta}{\theta^2}\ge0.
\label{eq:reviewentropyflux}
\end{equation}
The divergence of $\boldsymbol Q$ itself need not be nonnegative. This stress-power and heat-flux decomposition is consistent with other continuum nonequilibrium variational formulations \citep{GayBalmaz2017b}; its PDEL representation retains the same realized history in the weak donor and entropy receiver.

The continuum boundary identities in this appendix are analytical results. The Lean development treats the homogeneous scalar reduction and a separate fixed-box Eulerian MHD formulation; its scope is specified in Section~\ref{sec:formal}.


\FloatBarrier
\begingroup\small
\setlength{\bibsep}{2.5pt}
\interlinepenalty=10000
\bibliographystyle{unsrtnat}
\bibliography{references}
\endgroup
\end{document}